\documentclass[reqno,centertags]{amsart}
\usepackage{a4wide}
\usepackage{amssymb}
\usepackage{upref}
\newcommand{\bbN}{{\mathbb{N}}}
\newcommand{\bbR}{{\mathbb{R}}}

\newcommand{\bbZ}{{\mathbb{Z}}}
\newcommand{\bbC}{{\mathbb{C}}}

\newcommand{\calC}{{\mathcal C}}

\newcommand{\calD}{{\mathcal D}}

\newcommand{\calM}{{\mathcal M}}

\newcommand{\calF}{{\mathcal F}}
\newcommand{\calK}{{\mathcal K}}

\newcommand{\N}{\mathfrak{g}}
\newcommand{\g}{\mathfrak{g}}
\newcommand{\hatt}{\widehat}
\newcommand{\fnref}[1]{${}^{\ref{#1}}$}
\newcommand{\Pinf}{P_\infty}
\newcommand{\mini}{\wedge}
\newcommand{\maxi}{\vee}
\newcommand{\no}{\nonumber}
\newcommand{\lb}{\label}
\newcommand{\f}{\frac}
\newcommand{\uz}{{\underline{z}}}
\newcommand{\al}{\alpha}

\newcommand{\hatulmu}{\hat{\underline{\mu}}}

\newcommand{\pa}{\partial}

\newcommand{\tht}{\theta}
\newcommand{\bi}{\bibitem}
\newcommand{\ul}{\underline}

\newcommand{\ti}{\widetilde}

\renewcommand{\Im}{\text{\rm Im}}

\DeclareMathOperator{\KdV}{KdV}
\DeclareMathOperator{\AKNS}{AKNS}

\DeclareMathOperator{\sG}{sGmKdV}

\DeclareMathOperator{\TL}{Tl}
\newcommand{\Div}{\operatorname{Div}}

\newcommand{\bze}{{(0)}}
\newcommand{\uzero}{{\underline{0}}}

\newcommand{\uw}{{\underline{w}}}

\newcommand{\n}{\mathfrak{g}}
\allowdisplaybreaks
\numberwithin{equation}{section}

\newtheorem{theorem}{Theorem}[section]
\newtheorem{lemma}[theorem]{Lemma}
\newtheorem{corollary}[theorem]{Corollary}

\theoremstyle{definition}

\theoremstyle{remark}
\newtheorem{remark}[theorem]{Remark}

\newcommand{\abs}[1]{\lvert#1\rvert}

\begin{document}
\title[Dubrovin equations]{Dubrovin equations and
integrable systems on
hyperelliptic curves}
\author{Fritz Gesztesy}
\address{Department of Mathematics,
University of Missouri,
Columbia, MO 65211, USA}
\email{fritz@math.missouri.edu}
\urladdr{http://www.math.missouri.edu/people/faculty/fgesztesypt.html}
\author{Helge Holden}
\address{Department of Mathematical Sciences, Norwegian University of 
Science and Technology, N--7034 Trondheim, Norway}
\email{holden@math.ntnu.no}
\urladdr{http://www.math.ntnu.no/\~{}holden/}

\thanks{Research supported in part by the Research Council of Norway 
under grant
107510/410, the US National Science Foundation under grant DMS-9623121, and the
University of Missouri Research Board grant RB-97-086.}
\date{\today}
\subjclass{Primary 35Q53, 35Q55, 39A10, 58F07;
Secondary 35Q51, 35Q58}

\begin{abstract}
We introduce the
most general version of
Dubrovin-type
equations for divisors on a hyperelliptic curve
$\calK_\g$ of arbitrary
genus $\g\in\bbN$, and
provide a new argument for linearizing the corresponding
completely integrable flows. Detailed
applications to completely
integrable systems, including the KdV, AKNS, Toda,
and the combined sine-Gordon and mKdV
hierarchies, are made.
These
investigations uncover a new principle for $1+1$-dimensional
integrable
soliton equations
in the sense that the Dubrovin equations, combined with
appropriate trace
formulas, encode
all hierarchies of soliton equations associated with
hyperelliptic curves. In other words, completely
integable hierarchies of
soliton equations
determine Dubrovin equations and associated trace
formulas and, vice versa,
Dubrovin-type
equations combined with trace formulas permit the
construction of
hierarchies of
soliton equations.
\end{abstract}

\maketitle

\section{Introduction} \lb{s1}

The purpose of this paper is to re-examine Dubrovin
equations for divisors
on hyperelliptic
Riemann surfaces and to underscore their exceptional
role in connection
with completely
integrable hierarchies of soliton equations.

Starting from four
representative hierarchies,
the Korteweg--de Vries (KdV),
Ablowitz--Kaup--Newell--Segur (AKNS), Toda
lattice (Tl), and the combined sine-Gordon and mKdV
(sGmKdV) hierarchy, we derive a new argument
for linearizing the
corresponding completely integrable flows. As a result of these
investigations we show that a proper combination of Dubrovin equations
and trace formulas involving auxiliary divisors on hyperelliptic
curves encodes all information on the underlying completely integrable
hierarchy of soliton equations.

In Section \ref{s2} we briefly review some basic facts
on hyperelliptic
curves and
establish the
notation used throughout this paper. Section \ref{s3}
provides
a ``crash
course''
into the four
different hierarchies closely following the detailed
accounts in
\cite{BGHT98}, \cite{GH97},
\cite{GH98}, \cite{GR96}, and \cite{GRT96}. In
particular, we
outline an
elementary polynomial,
recursive approach to these hierarchies, as originally
developed by S.\ I.\ Al'ber
\cite{Al79}, \cite{Al81} in the KdV context,
and introduce the corresponding divisors on $\calK_\g$
associated with
their algebro-geometric
solutions. Section \ref{s4} relates the polynomial
recursion
relation
approach
with elementary symmetric
functions (and functions derived from them) of
$\mu_1 (x,t_r),
\dots ,\mu_\g (x,t_r)$, where $\mu_j(x,t_r)$ are
certain analogs of
Dirichlet-type
eigenvalues of the corresponding Lax operator. In
Section \ref{s5}
we study
Dubrovin equations and, based
on the results of Section \ref{s4}, provide a new
proof of the
fundamental fact
that a change of
coordinates effected by the Abel map straightens out
the flows on the
Jacobi variety
$J(\calK_\g)$ of $\calK_\g$. In Section \ref{s6} we
briefly recall the
theta function
representations of algebro-geometric solutions and
Section \ref{s7}
illustrates
our results by deriving
interesting connections between the $\KdV_\g$ and sG
equations, and AKNS and Toda hierarchies,
respectively. These
connections establish the
fundamental role played by the Dubrovin equations as
the common underlying
principle for
hierarchies of soliton equations. In particular, our
formalism establishes
an isomorphism between
the class of algebro-geometric solutions of these
integrable systems. Finally, Appendix \ref{A} collects
some useful results in
connection with Lagrange
interpolation formulas.

We emphasize that our results are not necessarily
restricted to
hyperelliptic curves. In
particular, the approach of this paper
applies to
Boussinesq-type curves using the polynomial recursion
formalism for the
Boussinesq
hierarchy developed in \cite{DGU98}, \cite{DGU99}.

Depending perhaps a bit on one's taste, the results of
this paper may at
first sight appear
somewhat discouraging as they clearly shift the emphasis
from individual
hierarchies of soliton
equations toward Dubrovin-type equations. On the positive side,
however, they
establish the Dubrovin
equations as a universal object underlying all hierarchies.

We note that our approach to completely integrable soliton equations
is close in spirit to that developed by M.\ S.\ Al'ber and S.\ I.\
Al'ber in a series of papers (see, e.g., \cite{Al79}--\cite{ALM97} and
the references therein).  While their approach focuses on
algebraically integrable systems and hence on a Hamiltonian formalism with
associated action and angle variables, our approach concentrates on
how a combination of elementary symmetric functions of
$\mu_1(x,t_1),\dots, \mu_\g(x,t_r)$ and certain trace formulas
generate completely integrable hierarchies of soliton equations and
their algebro-geometric solutions.

A different series of papers closely related to our investigations and
focusing on characterizing real-valued solutions of various soliton
equations was published by Chen, Chin, Lee, Neil, Ting, and Tracy
\cite{TCL84}--\cite{TTCL84}, \cite{TCL84a}--\cite{TNCC87}.  
These authors, however, appear to be
unaware of the prior work of M.\ S.\ Al'ber and S.\ J.\ Al'ber in this
field.

Finally, we stress that the use of elementary symmetric functions and
hence of trace formulas in terms of Dirichlet eigenvalues has a long
history in the context of integrable equations.  In fact, as early as
1975, Flaschka \cite{Fl75} characterized the real-valued periodic
potentials $q$ with finitely many stability intervals of the
associated Schr\"odinger operator $-d^2/dx^2+q$ in $L^2(\bbR;dx)$ as
stationary solutions of the KdV hierarchy using (regularized) trace
relations for Dirichlet eigenvalues associated with $q$ and the
underlying periodicity interval.

\section{Hyperelliptic curves} \lb{s2}

Fix $N\in\bbN_0$.  We briefly review hyperelliptic
Riemann surfaces
of the
type,
\begin{multline}
\calF_N(z,y)=y^2-R_{N+1}(z)=0, \quad R_{N+1}(z)=
\prod_{m=0}^{N}(z-E_m), \\
\quad \{E_m\}_{m=0,\dots,N}\subset\bbC, \quad E_m \neq E_{m'}
\text{  for
} m \neq m'.
\label{b1}
\end{multline}
The material of this section is standard and can be found, for
instance, in
\cite{FK92}.
The curve \eqref{b1} is compactified by adding one
point $P_\infty$ at infinity if $N$ is even, and two points
$P_{\infty_+}$
and  $P_{\infty_-}$
if $N$ is odd.

One introduces an appropriate set of\footnote{$\lfloor
x\rfloor=\sup\{y\in\bbZ \mid y\le
x\}$.}
$\lfloor N+1\rfloor/2$ nonintersecting cuts $\calC_j$ joining
$E_{m(j)}$ and $E_{m^\prime(j)}$ and $\calC_\infty$ joining
$E_{N}$ and
$\infty$ if $N$ is even.
Denote
\begin{equation}
\calC=\bigcup_{j\in J\cup\{\infty\}}\calC_j, \quad
\calC_j\cap\calC_k=\emptyset,
\quad j\neq k,\label{b2}
\end{equation}
where  $J\subseteq\{1,\dots,\lfloor N+1\rfloor/2 \}$. Define the
cut plane
\begin{equation}
\Pi=\bbC\setminus\calC, \label{b3}
\end{equation}
and introduce the holomorphic function
\begin{equation}
R_{N+1}(\cdot)^{1/2}\colon \Pi\to\bbC, \quad
z\mapsto \left(\prod_{m=0}^{N}(z-E_m) \right)^{1/2}\label{b4}
\end{equation}
on $\Pi$ with an appropriate choice of the square root branch in
\eqref{b4}.
Define
\begin{equation}
\calM_{\g}=\{(z,\sigma R_{N+1}(z)^{1/2}) \mid z\in\bbC, \;
\sigma\in\{\pm1\}
\}\cup\begin{cases}
\{P_\infty\} & \text{for $N$ even}, \\
\{P_{\infty_+},P_{\infty_-}\} & \text{for $N$ odd},
\end{cases}\label{b5}
\end{equation}
by extending $R_{N+1}(\cdot)^{1/2}$ to $\calC$. The hyperelliptic
curve
$\calK_\g$ is then the set
$\calM_{\g}$ with its natural complex structure obtained upon
gluing the
two sheets of $\calM_{\g}$
crosswise along the cuts. Finite points
$P$ on $\calK_\g$ are denoted by
$P=(z,y)$, where $y(P)$ denotes the meromorphic function on
$\calK_\g$
satisfying $\calF_N(z,y)=y^2-R_{N+1}(z)=0$;
$\calK_\g$ has  genus
$\g=\lfloor N+1\rfloor/2$.

A basis of $\g$ linearly
independent holomorphic differentials on $\calK_\g$ is given by
$z^{\ell-1}\,dz/y(P)$ for $\ell=1,\dots,\g$, and we introduce
\begin{equation}
d\omega_j (P)=\sum_{\ell=1}^\g c_{j,\ell}
\frac{z^{\ell-1}\, dz}{y(P)},\quad j=1,\dots,\g, \label{b6}
\end{equation}
with normalization,
\begin{equation}
\int_{a_k} d\omega_j=\delta_{j,k}, \label{b7}
\end{equation}
where $\{a_j,b_j\}_{j=1}^\g$ is a homology basis for $\calK_\g$.

Define the matrix $\tau=(\tau_{j,\ell})$ by
\begin{equation}
\tau_{j,\ell}=\int_{b_j}d\omega_\ell. \label{b8}
\end{equation}
Then $\Im(\tau)>0$ and $\tau_{j,\ell}=\tau_{\ell,j}$.

The Riemann theta function associated with $\calK_\g$ and the given
homology basis
$\{a_j,b_j\}_{j=1}^\g$, by definition, reads
\begin{equation}
\theta(\ul z)=\sum_{\ul n\in\bbZ^\g}\exp\big(2\pi
i(\ul n,\ul z)+\pi
i(\ul n,\tau \ul n)\big),
\quad \ul z\in\mathbb{C}^\g. \label{b9}
\end{equation}
We fix a base point $P_0$ on  $\calK_\g$ and define the
Abel map
$\underline{A}_{P_0}$ by
\begin{equation}
\underline{A}_{P_0}(P)=
\big(\int_{P_0}^P d\omega_1,\dots,\int_{P_0}^P d\omega_\g \big)
\pmod{L_\g}, \quad P\in\calK_\g, \label{b10}
\end{equation}
with period lattice
\begin{equation}
L_\g=\{\ul n+\tau\ul m \mid \ul n,\ul m\in\bbZ^\g  \}. \label{b11}
\end{equation}
Similarly, we introduce
\begin{equation}
\ul \al_{P_0}  \colon
\Div(\calK_\g) \to J(\calK_\g),\quad
\calD \mapsto \ul \al_{P_0} (\calD)
=\sum_{P \in \calK_\g} \calD (P) \ul A_{P_0} (P),
\label{aa47}
\end{equation}
where $\Div(\calK_\g)$ and $J(\calK_\g) = \bbC^\g/L_\g$ denote
the set of
divisors on $\calK_\g$ and
the Jacobi variety of $\calK_\g$, respectively.

In connection with divisors on $\calK_\g$ we shall employ the
following
(additive) notation,
\begin{multline}
\calD_{Q_0\ul Q}=\calD_{Q_0}+\calD_{\ul Q}, \quad \calD_{\ul
Q}=\calD_{Q_1}+\cdots +\calD_{Q_n},
\text{  etc.,}  \\
\text{for  }  {\ul Q}=(Q_1, \dots ,Q_n) \in \sigma^n \calK_\g,
\end{multline}
where for any $Q\in\calK_\g$,
\begin{equation}
\calD_Q \colon  \calK_\g \to\bbN_0, \quad
P \mapsto  \calD_Q (P)=
\begin{cases} 1 & \text{for $P=Q$},\\
0 & \text{for $P\in \calK_\g\setminus \{Q\}$}, \end{cases}
\end{equation}
and $\sigma^n \calK_\g$ denotes the $n$th symmetric product of
$\calK_\g$.

\section{The hierarchies} \lb{s3}
\setcounter{theorem}{0}

We give a brief presentation of the KdV, AKNS, sGmKdV,
and Toda
hierarchies based on a polynomial, recursive approach. The material
of this
section
originated with work of S.\ I.\ Al'ber \cite{Al79}, \cite{Al81} (see also
\cite{Al91}--\cite{ALM97}). It has been further
developed in
\cite{BGHT98}, \cite{GH97}, \cite{GH98}, \cite{GR96}, and
\cite{GRT96} and
we closely follow the
latter  sources. Common to all these hierarchies is that one
can naturally associate with each one of them a hyperelliptic
curve as
described in the
previous section.

\textbf{The KdV hierarchy.}  The Lax pair consists of a
second-order
linear
differential expression $L$  of Schr\"{o}dinger-type,
\begin{equation}
L(t_\g)=-\f{d^2}{dx^2}+V(x,t_\g), \quad t_\g \in \bbR\label{h1}
\end{equation}
and a differential expression $P_{2\g+1}(t_\g)$ of order $2\g+1$
defined
recursively as
follows.  Let $\{f_j\}_{j\in\bbN_0}$ be given by
\begin{align}
f_0 = 1, \quad f_{j,x} &= - \frac14 f_{j-1, xxx} + Vf_{j-1,x}
+ \frac12 V_x f_{j-1},\quad j\in\bbN. \label{h2}
\end{align}

Explicitly ($f_j=f_j(x,t_\g)$, $(x,t_\g)\in \bbR^2$),
\begin{equation}
f_1  = \tfrac12 V + c_1,   \quad
f_2= - \tfrac18 V_{xx} +\tfrac38 V^2 +
 c_1\tfrac12 V + c_2, \quad\text{etc.,} \label{h3}
\end{equation}
where $c_j\in \bbC$ are integration constants. Then one defines
\begin{equation}
P_{2\g+1}(t_\g)  = \sum_{j=0}^\g \big( f_j(t_\g)
\frac{d}{dx}-\frac12 f_{j,x}(t_\g) \big) L(t_\g)^{\g-j},
\quad \g\in\bbN_0,  \label{h4}
\end{equation}
and using the definition of $f_j$ in \eqref{h2}, one finds
that the
commutator of $L(t_\g)$ and
$P_{2\g+1}(t_\g)$ is in fact a multiplication operator.  Indeed,
the Lax
commutator representation
reads
\begin{equation}
L_{t_\g}(t_\g)-[P_{2\g+1}(t_\g),L(t_\g)]=V_{t_\g}-
2f_{\g+1,x}(x,t_\g) =
\KdV_\g(V) =0.
\label{laxpair}
\end{equation}
Explicitly, one obtains for the first few KdV equations,
\begin{align}
\KdV_0 (V) &= V_{t_0} - V_x =0, \no\\
\KdV_1 (V) & = V_{t_1} + \tfrac14 V_{xxx} -
\tfrac32 VV_x - c_1V_x=0,\\
\KdV_2 (V) &= V_{t_2} -\tfrac1{16} V_{xxxxx} +\tfrac58 VV_{xxx}
+\tfrac54 V_x V_{xx} \no\\
& \quad - \tfrac{15}8 V^2 V_x
 - c_2 V_x + c_1 \big(\tfrac14 V_{xxx} -\tfrac32 VV_x\big)=0,
\text{  etc.,} \no
\end{align}
where, of course, $\KdV_1 (V)$ is {\it the} KdV equation.

Next define the polynomial\footnote{The zeros
$\mu_j(x, t_\g)$ of $F_\g$
turn out to be
eigenvalues
associated with $L(t_\g)$ and a Dirichlet boundary condition
at the point
$x\in\bbR$. \lb{fn2}} of degree $\g$ in $z$
\begin{equation}
F_\g(z,x,t_\g)  = \sum_{j=0}^\g f_{\g-j}(x,t_\g)
z^j=\prod_{j=1}^\g(z-\mu_j(x,t_\g))
\lb{Fg}
\end{equation}
implying
\begin{equation}
-2V_{t_\g} = F_{\g,xxx} -4 (V-z) F_{\g,x} -2V_x F_\g.
\end{equation}
In the special stationary case, defined by $V_{t_\g} =0$,
this integrates to
\begin{equation}
\frac12 F_{\g,xx} F_\g - \frac14 F_{\g,x}^2 - (V-z) F_\g^2 =
R_{2\g+1}(z).
\end{equation}
Here $R_{2\g+1}$ is a monic polynomial of degree $2\g+1$
with zeros
$\{E_0,\dots, E_{2\g}\}$. Hence,
\begin{equation}
R_{2\g+1}(z) = \prod_{m=0}^{2\g} (z-E_m), \quad
\{E_m\}_{m=0,\dots,2\g}
\subset \bbC.
\end{equation}

The hyperelliptic curve
$\calK_\g$ is defined in terms of the stationary KdV hierarchy
obtained by
considering a
$t_\g$-independent function $V=V(x)$, resulting in
\begin{equation}
[P_{2\g+1},L]=2f_{\g+1,x}=0.
\end{equation}
The classical Burchnall--Chaundy theorem \cite{BC23},
\cite{BC28}, \cite{BC32} (see also \cite{Pr96}, \cite{Wi85})
states that
commuting
differential operators are algebraically related.  In the
present context
one finds
\begin{equation}
P_{2\g+1}^2=R_{2\g+1}(L)
\end{equation}
and thus the hyperelliptic curve $\calK_\g$ of
genus $\g$ is
of the type
$y^2=R_{2\g+1}(z)$, with
$N=2\g$ even when compared to Section \ref{s2}.

For later purpose we quote the following asymptotic high-energy
expansion\footnote{For an appropriate
choice of the sign of the square roots in \eqref{green}, the
left-hand side of
\eqref{green} equals the Green's function on the diagonal
(i.e., for
$x=x'$) of $L(t_\g)$.} (see
\cite{GD75}, \cite{GRT96}, \cite{Sc95})
\begin{equation}
\f{iF_\g(z,x,t_\g)}{2R_{2\g+1}(z)^{1/2}}
\underset{z\to i\infty}{=}
\f{i}{2z^{1/2}}\sum_{j=0}^\infty \hat f_j(x,t_\g)
z^{-j},
\label{green}
\end{equation}
where $\hat f_j$ denotes the homogeneous coefficients $f_j$
with all integration constants
equal to zero, $c_{\ell}=0, \,\ell\geq 1$, that is,
\begin{equation}
\hat f_0 = f_0 =1, \quad \hat f_j=f_j\vert_{c_{\ell}=0},
\quad \ell=1,\dots,j, \, j\in \bbN.
\end{equation}
We also introduce the following fundamental meromorphic
function on $\calK_\g$,
\begin{equation}
\phi (P,x,t_\g)  = \frac{iy(P)+\frac12 F_{\g,x} (z,x,t_\g)}
{F_\g (z, x,t_\g)}
= \frac{-H_{\g+1} (z,x,t_\g)}{iy(P) -\frac12 F_{\g,x}(z, x,t_\g)},
\quad (x,t_\g) \in\bbR^2
\lb{1.4.15}
\end{equation}
(the second equality in \eqref{1.4.15} serving as a definition
of the
polynomial $H_{\g+1}$ of
degree $\g+1$ with respect to $z$) and the time-dependent
Baker--Akhiezer
function
$\psi(P,x,x_0, t_\g, t_{0,\g})$ on
$\calK_\g \setminus \{ P_\infty\}$,
\begin{multline}
\psi (P,x,x_0, t_\g,t_{0,\g}) \\
=\exp \bigg( \int_{t_{0,\g}}^{t_\g}  ds\,
\big(  F_\g (z,x_0,s)
\phi (P,x_0,s)
-\tfrac12 F_{\g,x}(z,x_0, s)\big) + \int_{x_0}^x dx'\,
\phi(P,x',t_\g)\bigg),\\
(x,x_0, t_\g,t_{0,\g}) \in\bbR^4.
\lb{1.4.16}
\end{multline}
The divisor $(\phi(P,x,t_\g))$ of $\phi(P,x,t_\g)$ is
given by
\begin{equation}
(\phi(P,x,t_\g)) =
\calD_{{\hat {\nu}_0}(x,t_\g){\ul {\hat \nu}}(x,t_\g)} -
\calD_{P_{\infty}{\ul
{\hat \mu}}(x,t_\g)},
\end{equation}
where
\begin{multline}
{\ul {\hat \mu}}(x,t_\g) =
(\hat \mu_1(x,t_\g), \dots ,\hat \mu_\g(x,t_\g))
\in \sigma^\g \calK_\g,\\
\hat \mu_j(x,t_\g) = (\mu_j(x,t_\g), -\tfrac{1}{2}
F_{\g,x}(\mu_j(x,t_\g),x,t_\g)), \quad j=1, \dots,\g \lb{muk}
\end{multline}
denote the Dirichlet divisors\fnref{fn2} and
\begin{multline}
{\hat \nu}_0(x,t_\g){\ul {\hat \nu}}(x,t_\g) =
(\hat \nu_0(x,t_\g), \dots
,\hat \nu_\g(x,t_\g)) \in
\sigma^{\g+1}\calK_\g, \\
\hat \nu_{\ell}(x,t_\g) = (\nu_{\ell}(x,t_\g), \tfrac{1}{2}
F_{\g,x}(\nu_{\ell}(x,t_\g),x,t_\g)),
\quad  \ell=0,\dots, \g
\end{multline}
abbreviate the Neumann divisors\footnote{The zeros
$\nu_{\ell}(x,t_\g)$ of
$H_{\g+1}$ turn out to
be eigenvalues associated with $L(t_\g)$ and a Neumann
boundary condition
at the point
$x\in\bbR$.} derived from the zeros of
$H_{\g+1}(z,x,t_\g)$,
\begin{equation}
H_{\g+1}(z,x,t_\g) = \prod_{\ell =0}^\g (z-\nu_{\ell}(x,t_\g)).
\end{equation}
The importance of $\phi$ in connection with divisors
on hyperelliptic
curves was recognized by
Jacobi
\cite{Ja46} and applied to the KdV case by Mumford
\cite{Mu84}, Sect.
IIIa.1 and McKean \cite{Mc85}
(see also \cite{EF85}, \cite{Pr96}).

\textbf{The AKNS hierarchy.} The Lax pair consists of a
Dirac-type
matrix-valued
differential expression
\begin{equation}
M(t_\n) = i \begin{pmatrix} \f{d}{dx} & -q(x,t_\n) \\
p(x,t_\n) &
-\f{d}{dx}\end{pmatrix},
\quad t_\g \in \bbR, \label{21}
\end{equation}
and a matrix-valued differential operator $Q_{\n+1}(t_\g)$
of order $\n+1$.
To define
$Q_{\n+1}(t_\g)$ we proceed as follows.  Define
$\left\{f_{\ell}(x,t_\n)\right\}_{\ell\in\bbN_0}$,
$\left\{g_{\ell}(x,t_\n)\right\}_{\ell\in\bbN_0}$, and
$\left\{h_{\ell}(x,t_\n)\right\}_{\ell\in\bbN_0}$
recursively by
($(x,t_\g)\in\bbR^2$),
\begin{align}
 f_0&=-iq,& g_0&=1,& h_0&=ip,  \label{22}\\
 f_{\ell+1}&= \f{i}{2} f_{\ell,x} - i q g_{\ell+1},
 & g_{\ell+1,x}&= pf_{\ell}+ q h_{\ell},&
 h_{\ell+1}&= -\f{i}{2} h_{\ell,x} + i p g_{\ell+1},
 \quad \ell\in\bbN_0. \no
\end{align}
The $2\times 2$ matrix $Q_{\n+1}(t_\g)$ is then defined by
\begin{equation}
Q_{\n+1}(t_\n)=i \sum_{\ell=0}^{\n+1}\begin{pmatrix}
                  -g_{\n+1-\ell}(t_\g) & f_{\n-\ell}(t_\g) \\
-h_{\n-\ell}(t_\g) & g_{\n+1-\ell}(t_\g)
                   \end{pmatrix}M(t_\n)^{\ell},
\quad \n\in \bbN_0,\quad f_{-1}=h_{-1}=0, \label{24}
\end{equation}
and one verifies that the commutator of $Q_{\n+1}(t_\g)$
and $M(t_\g)$ becomes
\begin{equation}
[Q_{\n+1}(t_\n),M(t_\n)]= \begin{pmatrix}0
& -2if_{\n+1}(t_\g) \\ 2 i
h_{\n+1}(t_\g) & 0
\end{pmatrix},
\quad \n\in \bbN_0. \label{25}
\end{equation}
Consequently, the Lax commutator representation for the
AKNS hierarchy reads
\begin{multline}
\f{d}{dt_\n}M(t_\n) -
[Q_{\n+1}(t_\n),M(t_\n)]=\begin{pmatrix}p_{t_\n}(x,t_\n)-
2 h_{\n+1}(x,t_\n)\\q_{t_\n}(x,t_\n)- 2
f_{\n+1}(x,t_\n)\end{pmatrix}=\AKNS_\n(p,q)=0, \\
 \n\in \bbN_0.\label{223}
\end{multline}
The first few equations equal,
\begin{align}
\AKNS_0(p,q)&=\begin{pmatrix} p_{t_{0}}- p_{x} +
 c_1(-2 i p)  \\ q_{t_{0}} - q_x + c_1(2 i q) \end{pmatrix}
= 0, \no   \\
 \AKNS_1(p,q)&=\begin{pmatrix}
         p_{t_{1}}  +\f{i}{2} p_{_{xx}}-i p{^2}q +
    c_1\left( - p_{_x}\right)
    +c_2(-2 i p) \\[2mm]
      q_{t_{1}}-\f{i}{2} q_{xx}+i pq^2
    + c_1\left(- q_x\right)+c_2(2 i q)\end{pmatrix}=0,
\text{  etc.}\lb{226}
\end{align}
Next, define polynomials $F_\n$, $G_{\n+1}$, and $H_\n$
with respect
to $z\in \bbC$,
\begin{align}
F_\n(z,x,t_\g)&=
\sum_{\ell=0}^{\n}f_{\n-\ell}(x,t_\g)z^{\ell}
=-iq(x,t_\g)\prod_{j=1}^{\n}
(z-\mu_j(x,t_\g)),
\no \\ G_{\n+1}(z,x,t_\g)&=
\sum_{\ell=0}^{\n+1}g_{\n+1-\ell}(x,t_\g)z^\ell,
\label{29}\\
H_\n(z,x,t_\g)&=
\sum_{\ell=0}^{\n}h_{\n-\ell}(x,t_\g)z^\ell=
ip(x,t_\g)\prod_{j=1}^{\n}
(z-\nu_j(x,t_\g)).
\no
\end{align}
In the special stationary case, where $p_{t_\g}=q_{t_\g}=0$,
we infer from
the recursion
\eqref{22} that
\begin{equation}
\left( G_{\n+1}^2 - F_\n H_\n \right)_x =0
\label{213}
\end{equation}
and hence
\begin{equation}
G_{\n+1}^2- F_\n H_\n = R_{2\n+2}(z),
\label{214}
\end{equation}
where $R_{2\g+2}$ is a monic polynomial of degree $2\g+2$
with zeros
$\{E_0,\dots,E_{2\g+1}\}$. Thus,
\begin{equation}
R_{2\g+2}(z) = \prod_{m=0}^{2\g+1} (z-E_m),
\quad \{E_m\}_{m=0,\dots,2\g +1}
\subset \bbC.
\end{equation}

The stationary case determines the hyperelliptic curve
$\calK_\n$ of genus $\n$ of the type
$y^2=R_{2\N+1}(z)$, with $N=2\N+1$ when compared to
Section \ref{s2}. If $p=p(x)$ and
$q=q(x)$ are stationary solutions of the AKNS equation,
\begin{equation}
[Q_{\g+1},M]=0,  \text{  that is,  }  f_{\n+1}=h_{\n+1}=0,
\end{equation}
Burchnall--Chaundy's theorem implies that
\begin{equation}
Q_{\n+1}^2+R_{2\n+2}(M)=0.
\end{equation}

By studying the Green's matrix of $M$ one finds the
following asymptotic
high-energy expansion \cite{GR96},
\begin{align}
\f{F_\n(z,x,t_\g)}{R_{2\n+2}(z)^{1/2}}&
\underset{z\to i\infty}{=}
\f1z\sum_{k=0}^{\infty}\hat f_k(x,t_\g)z^{-k},
\lb{4.58} \\
\f{H_\n(z,x,t_\g)}{R_{2\n+2}(z)^{1/2}}&
\underset{z\to i\infty}{=}
\f1z\sum_{k=0}^{\infty}\hat h_k(x,t_\g)z^{-k}
\lb{4.58a}
\end{align}
for an appropriate determination of the square roots
in \eqref{4.58} and
\eqref{4.58a}. Here $\hat
f_j$ and $\hat h_j$ denote the homogeneous quantities
with vanishing
integration constants $c_{\ell}, \, \ell\geq 1$, that is,
\begin{align}
\hat f_0 = f_0 =-iq, \quad \hat f_j&=f_j\vert_{c_{\ell}=0},
\quad \ell=1,\dots,j, \, j\in \bbN, \lb{4.58b} \\
\hat h_0 = h_0 =ip, \, \, \, \, \quad \hat
h_j&=h_j\vert_{c_{\ell}=0}, \quad \ell=1,\dots,j, \, j\in \bbN.
\lb{4.58c}
\end{align}
We also record the functions
\begin{multline}
\phi(P,x,t_\n)=\f{y(P)+G_{\n+1}(z,x,t_\n)}{F_\n(z,x,t_\n)}=
     \f{-H_\n(z,x,t_\n)}{y(P)-G_{\n+1}(z,x,t_\n)},  \\
 P=(z,y) \in \calK_\n \label{415}
\end{multline}
and the Baker-Akhiezer vector,
\begin{align}
\Psi(P,x,x_0,t_\n,t_{0,\n})&=\begin{pmatrix}
             \psi_1(P,x,x_0,t_\n,t_{0,\n}) \\
              \psi_2(P,x,x_0,t_\n,t_{0,\n})
     \end{pmatrix}, \label{421} \\
\psi_1(P,x,x_0, t_\n,t_{0,\n}) &=
\exp \left( \int_{x_0}^x dx'\, (-iz+
q(x',t_\n)\phi(P,x',t_\n))  \right. \no \\
&\qquad
 \left. +i\int_{t_{0,\n}}^{t_\n} \, ds (F_\n (z,x_0,
s) \phi (P,x_0,s) -G_{\n+1} (z,x_0, s)) \right), \no \\
\psi_2(P,x,x_0,t_\n,t_{0,\n})&=
\phi(P,x,t_\n)\psi_1(P,x,x_0,t_\n,t_{0,\n}), \\
& \quad \quad \quad \quad \quad \quad \quad \quad P\in
\calK_\n\setminus
\{P_{\infty_{\pm}}\}, \,
(x,t_\n)\in\bbR^2.\no
\end{align}
The divisor of $\phi(P,x,t_\g)$ is given by
\begin{equation}
(\phi(P,x,t_\g)) = \calD_{P_{\infty_+}
{\ul {\hat \nu}}(x,t_\g)} -
\calD_{P_{\infty_-}{\ul
{\hat \mu}}(x,t_\g)},
\end{equation}
where
\begin{multline}
{\ul {\hat \mu}}(x,t_\g) =
(\hat \mu_1(x,t_\g), \dots ,\hat \mu_\g(x,t_\g))
\in \sigma^\g \calK_\g, \\
\hat \mu_j(x,t_\g) =
(\mu_j(x,t_\g), G_{\g+1}(\mu_j(x,t_\g),x,t_\g)), \quad
j=1,\dots, \g
\end{multline}
and
\begin{multline}
{\ul {\hat \nu}}(x,t_\g)
=(\hat \nu_1(x,t_\g), \dots ,\hat \nu_\g(x,t_\g))
\in \sigma^\g \calK_\g,\\
\hat \nu_j(x,t_\g) = (\nu_j(x,t_\g),
-G_{\g+1}(\nu_j(x,t_\g),x,t_\g)),
\quad j=1,\dots, \g.
\end{multline}

\textbf{The Toda hierarchy.} Let $(S^{\pm} f)(n)=
f^\pm(n)=f(n\pm1)$,
$n\in\bbZ$ denote the
shift operation on the lattice $\bbZ$. The equations in
the Toda hierarchy
are continuous
in time and discrete in the space variable. The Lax pair
consists of the
second-order difference
operator
\begin{equation}
L(t_\g) = a(t_\g) S^+ +a^-(t_\g) S^- -b(t_\g),
\quad t_\g\in\bbR \label{3.2.8}
\end{equation}
and a difference operator $P_{2\g+2}(t_\g)$ of order
$2\g+1$
\begin{multline}
 P_{2\g+2}(t_\g)=-L(t_\g)^{\g+1} +
\sum_{j=0}^\g ( g_j(t_\g)
+2a(t_\g) f_j(t_\g) S^+) L(t_\g)^{\g-j} +
f_{\g+1}(t_\g), \label{3.2.9} \\
 \g\in\bbN_0.
\end{multline}
Here $\{f_j(n,t_\g)\}_{j\in\bbN_0}$ and
$\{g_j(n,t_\g)\}_{j\in\bbN_0}$
satisfy the recursion
relations,
\begin{align}
\begin{split}
& f_0 = 1, \quad g_0 = -c_1,\\
& 2 f_{j+1} +g_j + g^-_j +2b
f_j =0, \quad j\in\bbN_0,\\
& g_{j+1} - g_{j+1}^- + 2\big(a^2 f^+_j -(a^-)^2f^-_j\big) +
b\big(g_j - g_j^-\big) =0, \quad j\in\bbN_0.
\end{split}\label{3.2.10}
\end{align}
The Lax commutator representation of the Toda hierarchy
then reads
\begin{multline}
 L_{t_\g}(t_\g) -[P_{2\g+2}(t_\g), L(t_\g)]
=\TL_\g (a,b)_1 S^+ -\TL_\g (a,b)_2+\TL_\g(a^-,b^-)_1 S^- =0,\\
\g\in\bbN_0, \label{3.2.11}
\end{multline}
where
\begin{align}
\begin{split}
\TL_\g (a, b)_1 & =  a_{t_\g} +a (g_\g^+
+g_\g  + f_{\g+1}^+  + f_{\g+1} + 2b^+ f_\g^+  )=0,
\label{3.2.12}\\
\TL_\g (a, b)_2 & =  b_{t_\g} +2 \big(b (g_\g +
f_{\g+1}) +a^2 f_\g^+
 -(a^-)^2 f^-_\g +b^2 f_\g\big)=0.
\end{split}
\end{align}
This is equivalent to
\begin{equation}
\TL_\g(a,b) =(\TL_\g (a,b)_1, \TL_\g (a,b)_2)^t =0,
\quad \g\in\bbN_0.
\label{3.2.13}
\end{equation}
The first few equations of the Toda hierarchy equal,
\begin{align}
\TL_0 (a,b) &=  \binom{a_{t_1} -a(b-b^+)}{b_{t_1}-
2\left((a^-)^2 -a^2
\right)} =0, \\[3mm]
\TL_1 (a,b) &=  \binom{a_{t_2} -a\left((a^+)^2 -(a^-)^2
+(b^+)^2 -b^2\right)}{b_{t_2} -2a^2 (b^+ +b) +2(a^-)^2 (b+b^-)}
 + c_1 \binom{ -a(b-b^+)}{ -2 \left((a^-)^2 -a^2\right)}
=0, \no \\
&\text{ etc.} \no
\end{align}
Next, define
\begin{align}
F_{\g} (z,n,t_\g) &=\sum_{j=0}^\g z^j f_{\g-j}(n,t_\g)=
\prod_{j=1}^\g
(z-\mu_j(n,t_\g)),\lb{3.2.21a}\\ G_{\g+1}(z,n,t_\g) &=-z^{\g+1} +
\sum_{j=0}^\g z^j g_{\g-j}(n,t_\g)
+ f_{\g+1}(n,t_\g).
\lb{3.2.21}
\end{align}
In the special stationary case, defined by $a_{t_\g}=b_{t_\g}=0$,
the
recursion formulas
\eqref{3.2.10} then imply
\begin{equation}
G_{\g+1}-4a^2 F_{\g}F_{\g}^+=G_{\g+1}^- -4(a^-)^2 F_{\g}^-
F_{\g}=R_{2\g+2}(z), \lb{poly1}
\end{equation}
where $R_{2\g+2}(z)$ is a lattice constant. By inspection,
$R_{2\g+2}(z)$
is a polynomial in $z$ of
degree $2\g+2$ with zeros $\{E_0,\dots,E_{2\g+1}\}$, that is,
\begin{equation}
R_{2\g+2}(z)=\prod_{m=0}^{2\g+1} (z-E_m),
\quad \{E_m\}_{m=0,\dots,2\g +1}
\subset \bbC.
\lb{poly2}
\end{equation}

Consider now the stationary hierarchy where $a=a(n)$ and
$b=b(n)$ satisfy
$[P_{2\g+2},L]=0$, or
\begin{align}
\begin{split}
g_\g^+ +g_\g  + f_{\g+1}^+  + f_{\g+1} + 2b^+ f_\g^+ =0,
\label{3.2.12aa}\\
b (g_\g + f_{\g+1}) +a^2 f_\g^+ -(a^-)^2 f^-_\g +b^2 f_\g=0.
\end{split}
\end{align}
Burchnall--Chaundy's theorem then states that
\begin{equation}
P_{2\g+2}^2=R_{2\g+2}(L). \lb{BC}
\end{equation}
Hence the hyperelliptic curve  $\calK_\g$ of genus
$\g$ reads
$y^2=R_{2\g+2}(z)$, and thus $N=2\g+1$
is odd in the terminology of Section \ref{s2}.

Studying the diagonal Green's function of $L(t_\g)$ yields
the high-energy
expansion \cite{BGHT98}
\begin{equation}
\f{F_\g(z,n,t_\g)}{R_{2\g+2}(z)^{1/2}}=
\sum_{j=0}^\infty \hat f_j(n,t_\g)
z^{-j-1} \text{ for
$\abs{z}> \Vert L \Vert$}, \lb{highen}
\end{equation}
with an appropriate choice of the radical in \eqref{highen}.
Here $\hat f_j$ and similarly $\hat g_j$, denote the
homogeneous coefficients $f_j$ and $g_j$ with vanishing
integration constants $c_{\ell}=0, \, \ell\geq 1$, that is,
\begin{align}
\hat f_0=1, \quad \hat f_j &=f_j\vert_{c_{\ell}=0},
\quad \ell=1,\dots,j, \, j\in\bbN, \\
\hat g_0=0, \quad \hat g_j &=g_j\vert_{c_{\ell}=0},
\quad \ell=1,\dots,j+1, \, j\in\bbN.
\end{align}

Furthermore, define the  meromorphic function
$\phi(P,n,t_\g)$ on $\calK_\g$ by
\begin{equation}
\phi(P,n,t_\g)  = \frac{-G_{\g+1}(z, n,t_\g)+
y(P)}{2a (n,t_\g)F_\g (z, n,t_\g)}
=\frac{-2a (n,t_\g) F_\g (z, n+1,t_\g)}
{ G_{\g+1} (z, n,t_\g) + y(P)}, \quad
P =(z, y)
\lb{3.3.20}
\end{equation}
using  relation \eqref{poly1}.
With the help of $\phi (P, n,t_\g)$ we define another
meromorphic function $\psi (P,n, n_0,t_\g,t_{0,\g})$
on $\calK_\g$, the
Baker--Akhiezer function, by
\begin{multline}
\psi (P,n,n_0,t_\g,t_{0,\g}) \\
=\exp \left( \int_{t_{0,\g}}^t \, ds \big(2a (n_0, s) F_\g
(z,n_0,s) \phi (P, n_0, s)
+G_{\g+1} (z,n_0, s)\big)\right) \times \\
\times \begin{cases}
\prod_{m=n_0}^{n-1} \phi (P,m,t_\g) & \text{for
$n \geq n_0+1$},\\
1 & \text{for $n=n_0$}, \\
\prod_{m=n}^{n_0-1} \phi (P,m,t_\g)^{-1} & \text{for
$n\leq n_0 -1$.}
\end{cases}\lb{3.6.8}
\end{multline}
The divisor $(\phi(P,n,t_\g))$ of $\phi(P,n,t_\g)$ is
given by
\begin{equation}
(\phi(P,n,t_\g)) =
\calD_{P_{\infty_+}{\ul {\hat \mu}}(n+1,t_\g)} -
\calD_{P_{\infty_-}{\ul
{\hat \mu}}(n,t_\g)},
\end{equation}
where
\begin{subequations} \lb{ddt}
\begin{align}
{\ul {\hat \mu}}(m,t_\g) &=
(\hat \mu_1(m,t_\g), \dots ,\hat \mu_\g(m,t_\g))
\in \sigma^\g \calK_\g,
 \quad m\in\bbZ, \lb{ddt1} \\
\hat \mu_j(n,t_\g) &=
(\mu_j(n,t_\g), -G_{\g+1}(\mu_j(n,t_\g),n,t_\g)),
\quad j=1,\dots,\g,
\lb{ddt2} \\
\hat \mu_j(n+1,t_\g) &= (\mu_j(n+1,t_\g),
G_{\g+1}(\mu_j(n+1,t_\g),n,t_\g)), \quad j=1,\dots,\g. \lb{ddt3}
\end{align}
\end{subequations}

\textbf{The sGmKdV hierarchy.} The combined sine-Gordon and
mKdV hierarchy is defined in terms
of a zero curvature formalism as follows. Introduce the
$2\times2$  matrices
\begin{align}
U(z,x,t_\N) & =-i\begin{pmatrix}\f12 u_x(x,t_\N) & 1 \\ z
& -\f12
u_x(x,t_\N)\end{pmatrix}, \quad (x,t_\N)\in\bbR^2,\lb{2.1} \\
\intertext{and}
V_\N(z,x,t_\N) &=\begin{pmatrix} -G_{\N-1}(z,x,t_\N)
& \f{1}{z}F_\N(z,x,t_\N) \\
H_\N(z,x,t_\N) & G_{\N-1}(z,x,t_\N)\end{pmatrix},
\quad  (x,t_\N)\in\bbR^2,\,  \N\in\bbN_0.
\lb{2.2}
\end{align}
Then the zero curvature relation reads
\begin{equation}
U_{t_\N}-V_{\N,x}+[U,V_\N]=0, \quad \N\in\bbN_0,\lb{2.33}
\end{equation}
resulting in the equations
\begin{subequations}\lb{2.34}
\begin{align}
u_{xt_\N}(x,t_\N)&=-2iG_{\N-1,x}(x,t_\N)-
2(H_\N(x,t_\N)- F_\N(x,t_\N)),
\lb{2.34a} \\
F_{\N,x}(x,t_\N)&=-i u_x(x,t_\N) F_\N(x,t_\N)-
2iz G_{\N-1}(x,t_\N),
\lb{2.34b} \\
H_{\N,x}(x,t_\N)&=i u_x(x,t_\N) H_\N(x,t_\N)+
2iz G_{\N-1}(x,t_\N).  \lb{2.34c}
\end{align}
\end{subequations}
Making the following polynomial ansatz
\begin{subequations}\lb{2.14}
\begin{align}
F_\N(z,x,t_\N)&=\sum_{j=0}^\N f_{\N-j}(x,t_\N) z^j=
\prod_{j=1}^\N(z-\mu_j(x,t_\N)),
\lb{2.14a}\\
H_\N(z,x,t_\N)&=\sum_{j=0}^\N h_{\N-j}(x,t_\N) z^j=
\prod_{j=1}^\N(z-\nu_j(x,t_\N)),
\lb{2.14b}\\
G_{-1}(z,x,t_\N)&=0,\quad G_{\N-1}(z,x,t_\N)
=\sum_{j=0}^{\N-1} g_{\N-1-j}(x,t_\N)z^j,\lb{2.14c}
\end{align}
\end{subequations}
one concludes\footnote{Observe that the
recursions \eqref{2.15a} and \eqref{2.15c}
are identical with the KdV recursion replacing $V$
by $w_\pm$. Hence explicit formulas for
the first few can be read off from \eqref{h3} upon
changing  $V$ into $w_\pm$.} (see
\cite{GH97} for a detailed discussion) ($f_j=f_j(x,t_\N)$, etc.)
\begin{subequations} \lb{2.15A}
\begin{align}
f_{\N}& =\alpha e^{-iu}, & h_{\N}  &=\beta e^{iu},
\quad \alpha,\beta\in\bbC,\, \N\in\bbN_0. \lb{2.15d}\\
f_0 & = 1, &
f_{j,x}&  = - \frac14 f_{j-1, xxx} + w_+ f_{j-1,x}
+ \frac12 w_{+,x} f_{j-1},
\quad j=1,\dots,\N, \, \N \in\bbN,  \lb{2.15a} \\
h_0 &  = 1, &
h_{j,x}& = - \frac14 h_{j-1, xxx} + w_- h_{j-1,x}
+ \frac12 w_{-,x} h_{j-1},  \lb{2.15c}
\quad j=1,\dots,\N, \, \N \in\bbN,
\end{align}
\end{subequations}
where
\begin{equation}
w_\pm=-\f14(u_x^2\pm2i u_{xx}), \lb{2.12}
\end{equation}
and
\begin{equation}
g_{-1}=0, \quad g_j=\f{i}{2}(f_{j,x}+i u_x f_{j})=
\f{i}{2}(-h_{j,x}+i u_xh_{j}),
\quad j=0,\dots,\N-1, \, \N\in\bbN.\lb{2.23}
\end{equation}
Explicitly,
\begin{equation}
g_0=-\tfrac12 u_x, \quad
g_1=\tfrac1{16} u_x^3+\tfrac18 u_{xxx}-
\tfrac{c_1}2 u_x, \text{  etc.}  \lb{2.25}
\end{equation}
We also list a few coefficients in the homogeneous case
where all integration constants $c_{\ell}, \, \ell\geq 1$
vanish,
\begin{align}
 \hat f_0&=f_0=1, &\hat g_0&=g_0=-\tfrac{1}{2} u_x,&
\hat h_0&=h_0=1, \quad \N\in\bbN, \no \\
 \hat{f}_j&=f_j\big|_{c_{\ell}=0}, &
\hat{g}_j&=g_j\big|_{c_{\ell}=0}, &
\hat{h}_j&=h_j\big|_{c_{\ell}=0}, \lb{2.17aa} \\
& & && \ell&=1,\dots,j, \, j=1,\dots,\N-1, \, \N \geq 2, \no \\
\hat f_\N&=f_\N=\alpha e^{-iu}, & \hat h_\N&=h_\N=\beta e^{iu}, &
 \N&\in\bbN_0. \lb{2.17aab}
\end{align}

\begin{remark} \lb{remark-sG}
The recursion for the sGmKdV hierarchy is anomalous
compared to the other hierarchies studied
in this paper.  For the KdV, AKNS as well as the Tl hierarchies
the functions $f_j$ (and $g_j$
and $h_j$ where applicable) are defined by the same recursion
formula for all $j\in\bbN$
irrespective of the given genus $\N$.  However, for the sG
hierarchy $f_\N$ and $h_\N$ are always
given by \eqref{2.15d} for $\g\in\bbN_0$.  This
raises a compatibility problem in the
recursion formalism. A proof of the solvability of the
recursion can be found in \cite{GH97}, Appendix C.
\end{remark}

The $\N$th sGmKdV equation is then
defined by
\begin{multline}
\sG_\N(u(x,t_\N))=u_{xt_\N}(x,t_\N)+2ig_{\N-1,x}(x,t_\N)
+2(\beta e^{iu(x,t_\N)}-\alpha e^{-iu(x,t_\N)})=0, \\
\N\in\bbN_0. \lb{2.36}
\end{multline}
Explicitly, the first few equations read
\begin{align}
\sG_0(u) & =  u_{xt_0}+2(\beta e^{iu}-\alpha e^{-iu})=0, \no \\
\sG_1(u) & =  u_{xt_1}-iu_{xx}+
2(\beta e^{iu}-\alpha e^{-iu})=0, \lb{2.37} \\
\sG_2(u) & =  u_{xt_2}+\tfrac{i}8(u_x^3+2u_{xxx})_x
-c_1 i u_{xx} +2(\beta e^{iu}-\alpha e^{-iu})=0,
\text{  etc.} \no
\end{align}
Observe that $\alpha=\beta=i/4$ and $\N=0$ yields the
well-known sine-Gordon equation in
light-cone coordinates.  Appropriate choices of $\alpha$
and $\beta$ in \eqref{2.36}
include the sinh-Gordon hierarchy, the corresponding elliptic
equations, the Liouville model,
as well as the modified KdV hierarchy (taking $\alpha=\beta=0$).

In the stationary case, where $u_{xt_\N}=0$, we find
\begin{equation}
\f{d}{dx}\bigg(zG_{\N-1}(z,x)^2+
F_\N(z,x) H_\N(z,x)\bigg)=0 \lb{2.5}
\end{equation}
and hence
\begin{equation}
zG_{\N-1}(z,x)^2+F_\N(z,x) H_\N(z,x)=P_{2\N}(z), \lb{2.6}
\end{equation}
where $P_{2\N}(z)$ is $x$-independent.  It  is more
convenient to define  $R_{2\N+1}(z)=zP_{2\N}(z)$ so that
\eqref{2.6} becomes
\begin{equation}
z^2G_{\N-1}(z,x)^2+zF_\N(z,x) H_\N(z,x)=R_{2\N+1}(z), \lb{2.7}
\end{equation}
where $R_{2\N+1}$ is a monic polynomial in $z$ of degree
$2\N+1$ of the form
\begin{equation}
R_{2\N+1}(z)=\prod_{m=0}^{2\N}(z-E_m),
\quad E_0=0,\, E_1,\dots,
E_{2\N}\in\bbC. \lb{3.1}
\end{equation}
This polynomial defines the hyperelliptic curve $\calK_\N$ of
 genus $\N$ by the relation
$y^2-R_{2\N+1}(z)=0$ and hence $N=2\N$ in the terminology of
Section \ref{s2}. $\calK_\N$ is compactified by
adding a  point $\Pinf$.

\begin{remark} \lb{remark-sG1a}
Observe that the sGmKdV-curve is a special case of the KdV
curve with the additional constraint $E_0=0$.
\end{remark}

\begin{remark} \lb{remark-sG2}
In the stationary case the choice of $\alpha$ and $\beta$ is
constrained by the relation
\begin{equation}
\alpha\beta=\prod_{j=1}^{2\N} E_j,
\end{equation}
as can be seen by inserting $z=0$ in \eqref{2.7}, using
\eqref{2.15d} and  \eqref{3.1}.
\end{remark}

We now return to the time-dependent formalism.
Let $\phi(P,x,t_\N)$ be the meromorphic function on
$\calK_\N$ given by
\begin{align}
&\phi(P,x,t_\N)=\f{y(P)-zG_{\N-1}(z,x,t_\N)}{F_\N(z,x,t_\N)}
=\f{zH_{\N}(z,x,t_\N)}{y(P)+zG_{\N-1}(z,x,t_\N)}, \lb{4.9} \\
& \hspace*{4cm} (x,t_\N)\in\bbR^2, \,
P=(z,y)\in\calK_\N\setminus\{\Pinf \}. \no
\end{align}
Hence the divisor $(\phi(P,x,t_\N))$ of $\phi(P,x,t_\N)$
reads
\begin{equation}
(\phi(P,x,t_\N))=\calD_{Q_0\ul\nu(x,t_\N)}-
\calD_{\Pinf\ul\mu(x,t_\N)}, \lb{4.10d}
\end{equation}
with
\begin{subequations}\lb{3.10e}
\begin{align}
\hat{\ul\mu}(x,t_\N)&=
(\hat\mu_1(x,t_\N),\dots,\hat\mu_\N(x,t_\N))
\in\sigma^\N\calK_\N, \lb{3.10e1}\\
\hat\mu_j(x,t_r)&=(\mu_j(x,t_r),-\mu_j(x,t_r)
G_{\N-1}(\mu_j(x,t_r),x,t_r))\in\calK_\N,
\quad j=1,\dots,\N, \no  \\
\hat{\ul\nu}(x,t_\N)&=
(\hat\nu_1(x,t_\N),\dots,\hat\nu_\N(x,t_\N))
\in\sigma^\N\calK_\N,\lb{3.10e2} \\
\hat\nu_j(x,t_r)&=(\nu_j(x,t_r),\nu_j(x,t_r)
G_{\N-1}(\nu_j(x,t_r),x,t_r))\in\calK_\N, \quad
j=1,\dots,\N.\no
\end{align}
\end{subequations}
The time-dependent Baker--Akhiezer function
\begin{equation}
\Psi(P,x,x_0,t_\N,t_{0,\N})=
\begin{pmatrix} \psi_1(P,x,x_0,t_\N,t_{0,\N}) \\
\psi_2(P,x,x_0,t_\N,t_{0,\N})
\end{pmatrix}\lb{4.11}
\end{equation}
is defined by
\begin{align}
&\psi_1(P,x,x_0,t_\N,t_{0,\N})=
\exp\bigg(-\f{i}2(u(x,t_\N)-u(x_0,t_{\N}))
\lb{4.12a}\\
&\qquad +i\int_{x_0}^x dx'\,\phi(P,x',t_\N)-
\int_{t_{0,\N}}^{t_\N}ds\,
\big(\f1z \ti F_\N(z,x_0,s)\phi(P,x_0,s)+
\ti G_{\N-1}(z,x_0,s) \big)\bigg),\no \\
&\psi_2(P,x,x_0,t_\N,t_{0,\N})=
-\psi_1(P,x,x_0,t_\N,t_{0,\N}) \phi(P,x,t_\N),
\lb{4.12b} \\
&\hspace*{6cm} P\in\calK_\N\setminus\{\Pinf\}, \,
(x,x_0,t_\N,t_{0,\N})\in\bbR^4. \no
\end{align}
Combining relations \eqref{2.14a}, \eqref{2.14b},
and \eqref{2.15d} one concludes
\begin{equation}
u(x,t_\N)=i\ln\bigg((-1)^\N\alpha^{-1}
\prod_{j=1}^\N\mu_j(x,t_\N)\bigg)
=-i\ln\bigg((-1)^\N\beta^{-1}
\prod_{j=1}^\N\nu_j(x,t_\N) \bigg). \lb{4.33}
\end{equation}

We will also need the following asymptotic high-energy expansion
\begin{equation}
\f{F_\N(z,x,t_\N)}{R_{2\N+1}(z)^{1/2}}\underset{z\to i \infty}{=}
\f{1}{z^{1/2}}\sum_{j=0}^\infty \hat f_j(x,t_\N)z^{-j}.
 \lb{high}
\end{equation}

%
%
\section{Symmetric functions} \lb{s4}
Let $\g\in\bbN$ be fixed and define
\begin{subequations} \lb{ind}
\begin{align}
\mathcal S_k&=\{\ul\ell=(\ell_1,\dots,\ell_k)\in\bbN^k \mid
\ell_1<\cdots<\ell_k\leq \g \}, \quad k\le \g,\lb{inds} \\
\mathcal I^{(j)}_k&=\{\ul\ell=
(\ell_1,\dots,\ell_k)\in\mathcal S_k\mid
\ell_m\neq j\}, \quad k\le \g-1.  \lb{indt}
\end{align}
\end{subequations}
Define
\begin{subequations} \lb{funct}
\begin{align}
\Psi_{0}(\ul\mu)&=1, \quad \Psi_{k}(\ul\mu)=
(-1)^k\sum_{\ul\ell\in\mathcal
S_k}\mu_{\ell_1}\cdots\mu_{\ell_k},
\quad k\le \g,\lb{functpsi} \\
\Phi_0^{(j)}(\ul\mu)&=1, \quad
\Phi_k^{(j)}(\ul\mu)=(-1)^k\sum_{\ul\ell\in\mathcal
I^{(j)}_k}\mu_{\ell_1}\cdots\mu_{\ell_k},
\quad k\le \g-1, \quad
\Phi_\g^{(j)}(\ul\mu)=0,
\lb{functphi}
\end{align}
\end{subequations}
where $\ul\mu=(\mu_{1},\dots,\mu_{\g})\in\bbC^\g$.
One recognizes the
simple pattern,
\begin{subequations} \lb{functs}
\begin{align}
\Psi_{1}(\ul\mu)&=-\sum_{\ell=1}^\g \mu_\ell,
\quad \Psi_{2}(\ul\mu)=
\sum_{\substack{\ell_1,\ell_2=1\\ \ell_1<\ell_2}}^\g
\mu_{\ell_1}\mu_{\ell_2}, \text{ etc.,} \lb{functspsi} \\
\Phi_1^{(j)}(\ul\mu)&=-
\sum_{\substack{\ell=1\\ \ell\neq j}}^\g \mu_\ell, \quad
\Phi_2^{(j)}(\ul\mu)=
\sum^\g_{\substack{\ell_1, \ell_2=1\\ \ell_1, \ell_2\neq j\\
\ell_1<\ell_2}}\mu_{\ell_1}
\mu_{\ell_2}, \text{ etc.} \lb{functsphi}
\end{align}
\end{subequations}
Let $E_0,\dots,E_N$ be $N+1$ complex numbers, where $N=2\g$ or
$N=2\g+1$ depending on the underlying hierarchy of soliton equations.  For
brevity we introduce
\begin{equation}
\ul E=(E_0,\dots,E_N).
\end{equation}
We will need the following elementary result.
\begin{lemma}\label{lemma1}
For $z\in\bbC$ such that $|z|>\max\{|E_0|,\dots, |E_N|\}$
we have
\begin{equation}
\bigg(\prod_{m=0}^N \big(1-\f{E_m}{z}\big)
\bigg)^{-1/2}=\sum_{k=0}^{\infty}c_k(\ul
E)z^{-k},
\end{equation}
where\footnote{$(2n-1)!!=1\cdot3\cdots (2n-1)$, and $(-1)!!=1$.}
\begin{equation}
c_0(\ul E)=1,\quad
c_k(\ul E)=\sum_{\substack{j_0,\dots,j_N=0\\
 j_0+\cdots+j_N=k}}^{k}
\f{(2j_0-1)!!\cdots(2j_N-1)!!}
{2^k j_0!\cdots j_N!}E_0^{j_0}\cdots E_N^{j_N},
\,\, k\in\bbN. \label{c}
\end{equation}
\end{lemma}
\begin{proof}
It suffices to apply the binomial expansion.
\end{proof}

The few first terms read
\begin{equation}
c_1(\ul E)=\f12\sum_{j=0}^N E_j, \quad
c_2(\ul E)=\f14\sum_{\substack{j,k=0\\ j\neq k}}^N
E_j E_k+\f38
\sum_{j=0}^N E_j^2, \text{ etc.}
\end{equation}
Next, assuming $\mu_j \neq \mu_{j'}$ for $j \neq j'$,
introduce the
$\g\times \g$ matrix
$U_\g(\ul\mu)$ by
\begin{equation}
U_1(\ul\mu)=1, \quad U_\g(\ul\mu)=\left(\frac{\mu_k^{j-1}}
{\prod_{m\neq k}^\g(\mu_k-\mu_m)} \right)_{j,k=1}^\g.
\end{equation}
\begin{lemma} \label{lemma2}
Suppose $\mu_j \neq \mu_{j'}$ for $j \neq j'$. Then
\begin{equation}
U_\g(\ul\mu)^{-1}=
\left(\Phi_{\g-k}^{(j)}(\ul\mu)\right)_{j,k=1}^\g.
\end{equation}
\end{lemma}
\begin{proof}
Observe that we may write
\begin{equation}
U_\g(\ul\mu)=\left(\frac{\mu_k^{j-1}}{F_\g^\prime(\mu_k)}
\right)_{j,k=1}^\g. \label{matr}
\end{equation}
Using Lagrange's interpolation result, Theorem
\ref{theoremA1} (replacing
$k$ by $\g-k$ in
\eqref{A1c}), proves the result.
\end{proof}

Of crucial importance for our approach is the fact that
we are able to
express $f_j$
and $F_r$ in terms of
elementary symmetric functions of $\mu_1,\dots,\mu_\g$.
The expression is
given below for the
homogeneous case only, denoted by $\hat f_j$ and $\hatt F_r$,
where the
integration constants
$c_\ell$ for $\ell\in\bbN$ vanish. We start with
$\hat f_j$.
\begin{lemma} \label{lemma3}
Let $c_j(\ul E)$ be defined as in \eqref{c}. Then we infer
the following
results for the KdV and
the Toda hierarchies\footnote{$n\mini m=\min\{n,m\}$.},
\begin{equation}
\hat f_j=\sum_{k=0}^{j\mini \g}
c_{j-k}(\ul E) \Psi_k(\ul \mu). \lb{f}
\end{equation}
For the AKNS hierarchy we obtain
\begin{equation}
\hat f_j=-iq\sum_{k=0}^{j\mini \g} c_{j-k}(\ul E)
\Psi_k(\ul \mu), \quad
\hat h_j=ip\sum_{k=0}^{j\mini \g} c_{j-k}(\ul E)
\Psi_k(\ul \nu), \lb{fAKNS}
\end{equation}
where $\ul \nu=(\nu_1,\dots,\nu_\g)$.\\
In the sGmKdV case we have
\begin{equation}
\hat f_j=\sum_{k=0}^{j} c_{j-k}(\ul E)
\Psi_k(\ul \mu), \quad
\hat h_j=\sum_{k=0}^{j} c_{j-k}(\ul E)
\Psi_k(\ul \nu), \quad
j=0,\dots,\g-1, \, \g\in\bbN, \lb{fsG}
\end{equation}
and
\begin{equation}
\hat f_\g= \Psi_\g(\ul \mu),  \quad \hat h_\g=
\Psi_\g(\ul \nu), \quad \g\in\bbN_0. \lb{fsG1}
\end{equation}
\end{lemma}
\begin{proof}  The proof is identical in all cases,
and is based on the
high-energy expansion of
the Green's function of the corresponding linear operator
$L$ or $M$ in the
Lax pair or zero curvature formulation of the
hierarchy considered. We provide the details for the KdV
hierarchy only.

Using Lemma \ref{lemma1} we find
\begin{align}
&\f{F_\g(z)}{R_{2\g+1}(z)^{1/2}}=
\f{\prod_{j=1}^\g(z-\mu_j)}{R_{2\g+1}(z)^{1/2}}
=\f{1}{z^{1/2}}\f{\prod_{j=1}^\g(1-\f{\mu_j}{z})}
{\prod_{m=0}^{2\g}(1-\f{E_m}{z
})^{1/2}}  \no\\
& \quad=\f{1}{z^{1/2}}\big(\sum_{j=0}^\g
\Psi_j(\ul\mu)z^{-j}\big)
\big(\sum_{m=0}^\infty c_m(\ul E)z^{-m}\big)
=\f{1}{z^{1/2}}\sum_{m=0}^\infty z^{-m}
\sum_{k=0}^{m\mini \g} c_{m-k}(\ul E) \Psi_k(\ul \mu). \no
\end{align}
Combining this result with the high-energy expansion
\eqref{green} proves
the result.
\end{proof}
\begin{remark}
Observe that the right-hand side of \eqref{f} is defined
for all
$x\in\bbR$, but when we sample
it at integer values $x=n\in\bbZ$, it coincides with
$\hat f_j(n)$ for the
Toda lattice.  Thus we
have in some sense a continuous extension of the Toda
hierarchy (cf. also
Lemma \ref{dubn}).
\end{remark}
\begin{theorem} \label{theorem1}
Let $r\in\bbN_0$. For both the KdV and the Tl case one
derives\footnote{$n\maxi m=\max\{n,m\}$.}
\begin{equation}
\hatt F_r(\mu_j)=\sum_{p=(r-\g)\maxi 0}^r c_p(\ul E)
\Phi_{r-p}^{(j)}(\ul\mu).
\end{equation}
For the AKNS hierarchy one infers
\begin{equation}
\hatt F_r(\mu_j)=-iq\sum_{p=(r-\g)\maxi 0}^r c_p(\ul E)
\Phi_{r-p}^{(j)}(\ul\mu), \quad
\hatt H_r(\nu_j)=ip\sum_{p=(r-\g)\maxi 0}^r c_p(\ul E)
\Phi_{r-p}^{(j)}(\ul\nu).
\end{equation}
For the sGmKdV hierarchy one concludes\footnote{Since
$r$ is independent of $\N$, one obtains $\hat f_r=\tilde
\alpha e^{-iu}$,
$\hat h_r=\tilde \beta e^{iu}$ with
$\tilde \alpha, \tilde \beta\in\bbC$ independent of
$\alpha, \beta$,
and $\hat f_q, \hat h_q, \, q=1,\dots,r-1$ constructed as in
\eqref{2.17aa}.}
\begin{align}
\f{\hatt F_r(\mu_j)}{\mu_j}&=
\sum_{p=(r-1-\N)\maxi 0}^{r-1} c_p(\ul E)
\Phi_{r-1-p}^{(j)}(\ul\mu)-\frac{\tilde
\alpha}{\alpha}\Phi_{\N-1}^{(j)}(\ul\mu), \no \\
\f{\hatt H_r(\nu_j)}{\nu_j}&=
\sum_{p=(r-1-\N)\maxi 0}^{r-1} c_p(\ul E)
\Phi_{r-1-p}^{(j)}(\ul\nu)-\frac{\tilde
\beta}{\beta}\Phi_{\N-1}^{(j)}(\ul\nu).\lb{sG-F}
\end{align}
\end{theorem}
\begin{proof}
It suffices to consider the KdV and sG cases. By definition
\begin{equation}
\hatt F_r(z)=\sum_{\ell=0}^r \hat f_{r-\ell} z^\ell=
\sum_{\ell=0}^r z^\ell
\sum_{m=0}^{(r-\ell)\mini \g}\Psi_m(\ul\mu) c_{r-\ell-m}(\ul E).
\end{equation}
Consider first the case  $r\le \g$. Then
\begin{equation}
\hatt F_r(z)=\sum_{p=0}^r c_p(\ul E) \sum_{\ell=0}^{r-p} z^\ell
\Psi_{r-\ell-p}(\ul\mu)
\end{equation}
and hence
\begin{equation}
\hatt F_r(\mu_j)=\sum_{p=0}^r c_p(\ul E)
\Phi_{r-p}^{(j)}(\ul\mu),
\end{equation}
using \eqref{A9}.  In the case when $r \geq \g+1$ we find
\begin{align}
\hatt F_r(z)&=\sum_{m=0}^\g \Psi_m(\ul E)
\sum_{p=0}^{r-m} z^{r-m-p} c_p(\ul
E) \no \\
&=\sum_{p=0}^{r-\g} c_p(\ul
E)\big(\sum_{\ell=0}^\g\Psi_\ell(\ul\mu)z^{\g-\ell}
\big)z^{r-\g-p}
+\sum_{p=r-\g+1}^r c_p(\ul E)\sum_{\ell=0}^{r-p}
\Psi_\ell(\ul\mu)
z^{r-p-\ell}\no \\
&=F_\g(z)\sum_{p=0}^{r-\g} c_p(\ul E)z^{r-\g-p}+
\sum_{p=r-\g+1}^r c_p(\ul E)
\sum_{\ell=0}^{r-p} \Psi_\ell(\ul\mu) z^{r-p-\ell}   \\
&=F_\g(z)\sum_{p=0}^{r-\g} c_p(\ul E)z^{r-\g-p}+
\sum_{p=r-\g+1}^r c_p(\ul E)
\sum_{m=0}^{r-p}\Psi_{r-p-m}(\ul\mu) z^m. \no
\end{align}
Hence
\begin{equation}
\hatt F_r(\mu_j)=\sum_{p=r-\g+1}^r c_p(\ul E)
\Phi_{r-p}^{(j)}(\ul\mu),
\end{equation}
by using \eqref{A9} again.

In the sGmKdV case we first observe the identity
\begin{equation}
\hatt F_r(z)=z \hatt F_{r-1}(z)+ \hat f_r,
\end{equation}
which implies
\begin{equation}
\f{\hatt F_r(\mu_j)}{\mu_j}=\hatt F_{r-1}(\mu_j)+
\f{\hat f_r}{\mu_j}=
\sum_{p=(r-1-\N)\maxi 0}^{r-1} c_p(\ul E)
\Phi_{r-1-p}^{(j)}(\ul\mu)-\frac{\tilde
\alpha}{\alpha}\Phi_{\N-1}^{(j)}(\ul\mu), \lb{symm}
\end{equation}
using $\tilde f_r=\tilde \alpha e^{-iu}$ and the trace
relation \eqref{4.33}.

\end{proof}
%
%
\section{Dubrovin equations and linearized flows}\lb{s5}
\setcounter{theorem}{0}

Dubrovin \cite{Du75} made the fundamental observation 
that the Dirichlet divisors for the KdV equation satisfy
a first-order
system of differential
equations.  Solving this system can then be used to
recover the function
$V$ by appealing to a
trace formula (cf.\ \eqref{trace}).

Before we state the Dubrovin equations we need some
notation. Let
$\g\in\bbN$. We start by
constructing the hierarchies as explained in Section
\ref{s3}.  In
particular, we
construct the function
$F_\g$ with its zeros $\ul\mu=(\mu_1,\dots,\mu_\g)$,
and define the
corresponding hyperelliptic
curve $\calK_\g$.  (In the AKNS case we also construct
the function
$H_\g$.) Next, fix an $r\in\bbN_0$, and construct the
function $F_r$.  The integration
constants in the definition of $F_r$ are assumed to be independent of
those used to
construct $F_\g$, and to
emphasize this fact we denote it by $\ti F_r$ and the
corresponding constants by
$\tilde c_\ell$.  The
Dubrovin equations give the evolution of
$\ul\mu=(\mu_1,\dots,\mu_\g)$ in
terms of the deformation (time)
parameter $t_r$ according to the $r$th equation in the hierarchy
considered.

{\bf The KdV hierarchy.} In
our setting the Dubrovin equations for the KdV hierarchy
read \cite{DT76a},
\cite{Di91}, Sect.\ 12.3,
\cite{GD79}, \cite{GRT96}, \cite{Le87}, Chs.\ 10, 12,
\cite{Ma86}, Ch.\ 4,
\begin{subequations}\label{dubk}
\begin{align}
\frac{\partial}{\partial x}\mu_{j}(x,t_r)&=-2i\,
\frac{y(\hat \mu_j(x,t_r))}
{\prod_{\ell\neq j}^\g(\mu_j(x,t_r)-
\mu_\ell(x,t_r))},\lb{dubkx}\\
\frac{\partial}{\partial
t_r}\mu_{j}(x,t_r)&=
\ti F_r(\mu_j(x,t_r),x,t_r)\frac{\partial}{\partial
x}\mu_{j}(x,t_r)\label{dubkt}\\
&=-2i\,\frac{y(\hat \mu_j(x,t_r))}
{\prod_{\ell\neq j}^\g(\mu_j(x,t_r)-\mu_\ell(x,t_r))}\ti
F_r(\mu_j(x,t_r),x,t_r), \no
\end{align}
\end{subequations}
for $j=1,\dots,\g$. The initial data for
\eqref{dubk} on $\calK_\g$ equal
\begin{equation}
\hatulmu(x,t_{0,r})=\hatulmu^{(0)}(x),
\end{equation}
where $\hatulmu=(\hat\mu_1,\dots, \hat\mu_\n)$ denotes
\begin{equation}
\hat\mu_j(x,t_r)=(\mu_j(x,t_r),-
\f{i}{2}F_{\n,x}(\mu_j(x),x,t_r))\in\calK_\n,
\quad j=1,\dots,\n.
\end{equation}
We remark that \eqref{dubkx} is an immediate consequence
of \eqref{Fg} and
\eqref{muk}, while
\eqref{dubkt} follows from \eqref{muk} and
\begin{equation}
F_{\g,t_r} = \ti F_r F_{\g,x} -\ti F_{r,x} F_\g
\end{equation}
upon taking $z=\mu_j(x,t_r)$.

{\bf The AKNS hierarchy.} In this case the Dubrovin
equations for
$\hatulmu=(\hat\mu_1,\dots, \hat\mu_\n)$ are given
by \cite{CJ87}, \cite{GR96},
\begin{subequations}\label{akns}
\begin{align}
\frac{\partial}{\partial x}\mu_{j}(x,t_r)&=-2i\,
\frac{y(\hat \mu_j(x,t_r))}
{\prod_{\ell\neq j}^\g(\mu_j(x,t_r)-\mu_\ell(x,t_r))},
\lb{aknsx}\\
\frac{\partial}{\partial
t_r}\mu_{j}(x,t_r)&=-\f{\ti
F_r(\mu_j(x,t_r),x,t_r)}{iq(x,t_r)}\frac{\partial}
{\partial
x}\mu_{j}(x,t_r)\label{aknst}\\
 &=2\,\frac{y(\hat \mu_j(x,t_r))}
{q(x,t_r)\prod_{\ell\neq j}^\g(\mu_j(x,t_r)-\mu_\ell(x,t_r))}\ti
F_r(\mu_j(x,t_r),x,t_r), \no
\end{align}
\end{subequations}
for $j=1,\dots,\g$, with initial data on $\calK_\g$
\begin{equation}
\hatulmu(x,t_{0,r})=\hatulmu^{(0)}(x),
\end{equation}
where
\begin{equation}
\hat\mu_j(x,t_r)=(\mu_j(x,t_r),
G_{\n+1}(\mu_j(x),x,t_r))\in\calK_\n,
\quad j=1,\dots,\n.
\end{equation}

For the corresponding evolution of
$\ul {\hat \nu}=(\hat \nu_1,\dots,\hat
\nu_\g)$ we have
\begin{subequations} \label{nu}
\begin{align}
&\frac{\partial}{\partial x}\nu_{j}(x,t_r)=-2i\,
\frac{y(\hat \nu_j(x,t_r))}
{\prod_{\ell\neq j}^\g(\nu_j(x,t_r)-
\nu_\ell(x,t_r))},\label{nux} \\
&\frac{\partial}{\partial
t_r}\nu_{j}(x,t_r)=
\ti H_r(\nu_j(x,t_r),x,t_r)\frac{\partial}{\partial
x}\nu_{j}(x,t_r)\label{nut}\\
&\quad =-2\,\frac{y(\hat \nu_j(x,t_r))}
{p(x,t_r)\prod_{\ell\neq j}^\g(\nu_j(x,t_r)-\nu_\ell(x,t_r))}\ti
H_r(\nu_j(x,t_r),x,t_r), \no
\end{align}
\end{subequations}
for $j=1,\dots,\g$, with initial data on $\calK_\g$
\begin{equation}
\ul {\hat \nu}(x,t_{0,r})=\ul {\hat \nu}^{(0)}(x),
\end{equation}
where
\begin{equation}
\hat\nu_j(x,t_r)=(\nu_j(x,t_r),
-G_{\n+1}(\nu_j(x),x,t_r))\in\calK_\n,
\quad j=1,\dots,\n.
\end{equation}

{\bf The Toda hierarchy.} Here the Dubrovin equations for
$\hatulmu=(\hat\mu_1,\dots,
\hat\mu_\n)$ read \cite{BGHT98}, \cite{DT76b},
\cite{Mo76}, \cite{To89}, Ch.4,
\begin{equation}
\frac{\partial}{\partial t_r}\mu_{j}(n,t_r)
 =2\,\frac{y(\hat \mu_j(n,t_r))}
{\prod_{\ell\neq j}^\g(\mu_j(n,t_r)-\mu_\ell(n,t_r))}\ti
F_r(\mu_j(n,t_r),n,t_r),
\lb{dubt}
\end{equation}
for $j=1,\dots,\g$, with initial data on $\calK_\g$
\begin{equation}
\hatulmu(n,t_{0,r})=\hatulmu^{(0)}(n),
\end{equation}
where
\begin{equation}
\hat\mu_j(n,t_r)=(\mu_j(n,t_r),
-G_{\n+1}(\mu_j(n),n,t_r))\in\calK_\n,
\quad j=1,\dots,\n. \lb{td}
\end{equation}

We note that \eqref{aknsx} and \eqref{nux} formally
coincide with
\eqref{dubkx}. The case of the
Toda hierarchy, however, is quite different since
\eqref{dubt} concerns the
$t_r$-dependence of $\ul
{\hat \mu} (n,t_r)$ and no analogous first-order
nonlinear difference
equation concerning the
$n$-dependence of $\ul {\hat \mu} (n,t_r)$ (i.e.,
an analog of
\eqref{dubkx} or \eqref{aknsx})
appears to be known. In this context we refer the
reader to Lemma
\ref{dubn}, where we continue this
discussion.

{\bf The sGmKdV hierarchy.} Finally, in the case
of the sGmKdV hierarchy
the equations for $\ul\mu$ read \cite{GH97}
\begin{align}
\mu_{j,x}(x,t_r)&=-2i\f{y(\hat\mu_j(x,t_r))}
{\prod_{\substack{\ell=1\\ \ell\neq j}}^\N(\mu_j(x,t_r)-
\mu_\ell(x,t_r))},
\lb{4.26}\\
\mu_{j,t_r}(x,t_r)&=
2\f{\ti F_r(\mu_j(x,t_r),x,t_r)}{\mu_j(x,t_r)}
\f{y(\hat\mu_j(x,t_r))}{\prod_{\substack{\ell=1\\ \ell\neq
j}}^\N(\mu_j(x,t_r)-\mu_\ell(x,t_r))},
\lb{4.26a} \\
& \hspace*{5cm} j=1, \dots, \N, \, (x,t_r)\in\Omega, \no
\end{align}
with initial data
\begin{equation}
\hat\mu_j(x_0,t_{0,r})\in\calK_\N,
\quad j=1, \dots, \N. \lb{4.27a}
\end{equation}
The corresponding equations for $\ul\nu$ equal
\begin{align}
\nu_{j,x}(x,t_r)&=-2i\f{y(\hat\nu_j(x,t_r))}
{\prod_{\substack{\ell=1\\ \ell\neq j}}^\N(\nu_j(x,t_r)-
\nu_\ell(x,t_r))},
\lb{4.29} \\
\nu_{j,t_r}(x,t_r)&=2\f{\ti H_r(\nu_j(x,t_r),x,t_r)}
{\nu_j(x,t_r)}
\f{y(\hat\nu_j(x,t_r))}{\prod_{\substack{\ell=1\\ \ell\neq
j}}^\N(\nu_j(x,t_r)-\nu_\ell(x,t_r))},
\lb{4.29a} \\
& \hspace*{5cm} j=1, \dots, \N, \, (x,t_r)\in\Omega, \no
\end{align}
with initial conditions
\begin{equation}
\hat\nu_j(x_0,t_{0,r})\in\calK_\N,
\quad j=1, \dots, \N. \lb{4.27b}
\end{equation}

Next we will prove that the Abel map provides a clever
change of
coordinates that linearizes the
Dubrovin flows. This will turn out to be a consequence
of the fact that
$\ti F_r(\mu_j)$ can be
expressed as a linear combination of the functions
$\Phi_k^{(j)}$.  Using
Theorem
\ref{theorem1} it is immediate that this is not only
the case for the KdV
hierarchy, but also for
all the other hierarchies discussed in this paper.

\begin{theorem} \label{mainthm}
Suppose $\ul\mu (x,t_r)=
(\mu_1 (x,t_r),\dots,\mu_\g (x,t_r))$ satisfies the
Dubrovin equations
\eqref{dubk} and assume that $\mu_j \neq \mu_{j'}$ for
$j \neq j'$. Let
$r\in\bbN_0$ and introduce
\begin{equation}
\ti F_r(\mu_j)=\sum_{k=0}^{r\mini \g} d_{r,k}
\Phi_k^{(j)}(\ul\mu), \quad
d_{r,0},\dots,d_{r,r\mini \g}\in\bbC.
\label{assump}
\end{equation}
Then the Abel map
\begin{equation}
\underline{A}_{P_0}(\hat\mu_j(x,t_r))=
(\underline{A}_{P_0,1}(\hat\mu_j(x,t_r)),
\dots,
\underline{A}_{P_0,\g}(\hat\mu_j(x,t_r)))
\end{equation}
linearizes the Dubrovin flows \eqref{dubk} in
the sense that
\begin{equation}
\frac{\partial}{\partial t_r}
\sum_{j=1}^\g\underline{A}_{P_0,k}
(\hat\mu_j(x,t_r))=-2i \sum_{\ell=1\maxi (\N-r)}^{\g}
c_{k,\ell} d_{r,\N-\ell}
\lb{linear}
\end{equation}
and hence\footnote{The situation here resembles the one
in classical
mechanics where, by a  canonical change
to cyclic coordinates, the momentum
$p_j$ becomes a constant of motion and thus $q_j(t)=
q_j(t_0)+p_j (t-t_0)$
is linear in time.\lb{fn8}}
\begin{equation}
{\ul \alpha}_{P_0}(\calD_{\ul {\hat \mu} (x,t_r)}) 
= {\ul
\alpha}_{P_0}(\calD_{\ul {\hat \mu}
(x_0,t_{0,r})})  -2i (x-x_0)c_{k,\g} d_{0,0}
-2i (t_r -t_{0,r}) \sum_{\ell =1\maxi (\N-r)}^{\g} c_{k,\ell}
d_{r,\N-\ell}. \lb{alpha}
\end{equation}
\end{theorem}
\begin{proof}
One computes,
\begin{align}
&\frac{\partial}{\partial t_r}
\sum_{j=1}^\g\underline{A}_{P_0,k}
(\hat\mu_j(x,t_r))=
\frac{\partial}{\partial t_r}
\sum_{j=1}^\g\int_{P_0}^{\hat \mu_j(x,t_r)}\,
d\omega_k \no \\
&=\frac{\partial}{\partial t_r}
\sum_{j=1}^\g \sum_{\ell=1}^\g c_{k,\ell}
\int_{P_0}^{\hat\mu_j (x,t_r)}\frac{z^{\ell-1}\, dz}
{y(P)} \no\\
& = \sum_{j=1}^\g \sum_{\ell=1}^\g c_{k,\ell}
\frac{\mu_j (x,t_r)^{\ell-1}}{y(\hat \mu_j (x,t_r))}
\frac{\partial}{\partial t_r}
\mu_{j} (x,t_r)\no \\
& = -2i\sum_{j=1}^\g \sum_{\ell=1}^\g
c_{k,\ell}\frac{\mu_j (x,t_r)^{\ell-1}}
{y(\hat \mu_j (x,t_r))}\frac{y(\hat
\mu_j (x,t_r))}
{\prod_{m\neq j}^\g(\mu_j (x,t_r)-\mu_m (x,t_r))}
\ti F_r(\mu_j (x,t_r)) \no\\
& = -2i\sum_{j=1}^\g \sum_{\ell=1}^\g
c_{k,\ell} U_\g(\ul\mu (x,t_r))_{\ell,j}
\ti F_r(\mu_j (x,t_r)) = -2i
\sum_{\ell=1\maxi (\N-r)}^{\g}
c_{k,\ell} d_{r,\N-\ell}, \lb{mt}
\end{align}
using Lemma \ref{lemma2} in the final step.  As for the  $x$-variation,
we observe that $t_0$-derivative of
$\mu_j$ coincides with the $x$-derivative in \eqref{dubk}, and hence it is a
special case of \eqref{mt}.  This proves the theorem.
\end{proof}

\begin{corollary}
The Abel map linearizes the Dubrovin flows for the
KdV, AKNS, Tl, as well as the sGmKdV hierarchies.
\end{corollary}
\begin{proof}
Theorem \ref{theorem1} shows that $\ti F_r(\mu_j)$
(and $\ti H_r(\nu_j)$ in
the AKNS case) indeed satisfies the assumption \eqref{assump} of
Theorem \ref{mainthm}, and hence the key calculation \eqref{mt} carries over
to the AKNS, Tl, and sGmKdV systems.  The special case $r=0$ gives
the
$x$-variation in all but the sGmKdV case which, however can easily
be verified by explicit computation.
\end{proof}

\begin{remark}
We provide a few more details in the AKNS case.
Suppose $\ul\mu (x,t_r)$
satisfies \eqref{akns}
and similarly,
$\ul\nu (x,t_r)$
satifies
\eqref{nu}, with $\mu_j \neq
\mu_{j'}$ and
$\nu_j \neq \nu_{j'}$ for $j \neq j'$. Let $r\in\bbN_0$ and
introduce
\begin{equation}
\ti F_r(\mu_j)=-iq\sum_{k=0}^{r\mini \g} d_{r,k}
\Phi_k^{(j)}(\ul\mu), \quad
\ti H_r(\nu_j)=ip\sum_{k=0}^{r\mini \g} e_{r,k}
\Phi_k^{(j)}(\ul\nu).
\end{equation}
Then \eqref{linear} and \eqref{alpha} hold. In
addition, one obtains the following results for the analog of Neumann
divisors $\ul\nu(x,t_r)$.
\begin{equation}
\frac{\partial}{\partial t_r}
\sum_{j=1}^\g\underline{A}_{P_0,k}
(\hat\nu_j(x,t_r))=-2i \sum_{\ell=1\maxi (\N-r)}^{\g}
c_{k,\ell} e_{r,\N-\ell}
\lb{linear_nu}
\end{equation}
and hence
\begin{equation}
{\ul \alpha}_{P_0}(\calD_{\ul {\hat \nu} (x,t_r)})
= {\ul
\alpha}_{P_0}(\calD_{\ul {\hat \nu}
(x_0,t_{0,r})})  -2i (x-x_0) c_{k,\g} e_{0,0}
-2i (t_r -t_{0,r}) \sum_{\ell =1\maxi (\N-r)}^{\g} c_{k,\ell}
e_{r,\N-\ell}. \lb{beta}
\end{equation}
\end{remark}

Necessary and sufficient conditions on Lax pairs to linearize the flow
$t\to L_t$ on $J(C)$, where $\{L_t\}$ represents a dymamical system on
the Jacobi variety $J(C)$, with $C$ the underlying spectral curve,
have been considered by Griffiths \cite{Gr85}.  While he considers Lax
equations within a cohomological framework, our approach is much more
modest in scope but in turn reduces the linearization problem to an elementary
exercise in symmetric functions.

Solving these equations we can recover the solution of the
integrable equation using {\it trace formulas}. For the KdV
hierarchy we have the classical trace formula
\begin{equation}
V(x,t_r)=\sum_{m=0}^{2\g} E_m -2\sum_{j=1}^\g \mu_j(x,t_r).
\lb{trace}
\end{equation}
For the
AKNS hierarchy we have
\begin{subequations} \lb{aktr}
\begin{align}
\f{p_x(x,t_r)}{p(x,t_r)}
&=i\sum_{m=0}^{2\g+1}E_m-
2i\sum_{j=1}^{\n}\nu_{j}(x,t_r),\lb{aktrp} \\
\frac{q_x(x,t_r)}{q(x,t_r)}&=-i\sum_{m=0}^{2\g+1}E_m
+2i\sum_{j=1}^\g\mu_j(x,t_r), \lb{aktrq}
\end{align}
\end{subequations}
while for the Toda hierarchy one obtains
\begin{subequations} \lb{ttr}
\begin{align}
a(n,t_r)^2&=
-\frac12\sum_{j=1}^\g \frac{G_{\g+1}(\mu_j(n,t_r),n,t_r)}
{\prod_{k\neq j}^\g(\mu_j(n,t_r)-\mu_k(n,t_r))} \no \\
&-\frac14 b(n,t_r)^2-\frac14\sum_{j=1}^\g
\mu_j(n,t_r)^2+\frac18\sum_{m=0}^{2\g+1}E_m^2,
\lb{ttra} \\
b(n,t_r)&=-\frac12\sum_{m=0}^{2\g+1}E_m+
\sum_{j=1}^\g\mu_j(n,t_r). \lb{ttrb}
\end{align}
\end{subequations}
The ``trace'' relation for the sGmKdV hierarchy (it would be more
appropriate to call this a ``determinant'' relation) was given in
\eqref{4.33},
\begin{equation}
u(x,t_\N)=i\ln\bigg((-1)^\N\alpha^{-1}
\prod_{j=1}^\N\mu_j(x,t_\N)\bigg)
=-i\ln\bigg((-1)^\N\beta^{-1}
\prod_{j=1}^\N\nu_j(x,t_\N) \bigg). \lb{4.33aa}
\end{equation}

\begin{remark}\lb{remark1}
It is important to observe  that if one {\it postulates}
the Dubrovin equations
\eqref{dubk}, and {\it defines} $V$ using the trace
formula \eqref{trace},
one could show by
a long and tedious calculation that $V$ indeed satisfies
the $r$th KdV
equation with the
correct initial condition.  The same applies, of course,
to the AKNS, Toda, and sGmKdV hierarchies.
\end{remark}

\begin{remark}
For simplicity we assumed $\mu_j(x,t_r)
\neq \mu_{j'}(x,t_r)$ for $j \neq
j'$ in Theorem
\ref{mainthm}. In the self-adjoint cases, where
$\{E_m\}_{m=0,\dots,N}
\subset \bbR$, this
condition is automatically fulfilled for all
$(x,t_r)\in\bbR^2$ since then
all $\mu_j(x,t_r)$ are
separated from each other by spectral gaps of
$L(t_r)$ or $M(t_r)$. In the
general nonself-adjoint
case this is no longer true and collisions between
the $\mu_j$'s become
possible. Nevertheless the
Dubrovin equations, properly desingularized near such
collision points,
stay meaningful as
demonstrated in detail by Birnir \cite{Bi86a},
\cite{Bi86b} in the case of
complex-valued KdV solutions. In particular,
\eqref{alpha} (and \eqref{beta}) remain valid in the
presence of
such collisions due to the continuity of
${\ul \alpha}_{P_0}(\cdot)$.
\end{remark}

We already mentioned in the paragraph following \eqref{td}
that the
Dubrovin equations for the
Toda hierarchy differ from the ones associated with the
remaining soliton
hierarchies in the sense
that they do not seem to govern the $n$-dependence of
$\ul {\hat \mu}
(n,t_r)$. We now show how to
use Theorem
\ref{mainthm} to obtain a first-order Dubrovin system
for $\ul {\hat \mu}
(x,t_r)$ in $x$, whose
solution coincides with $\ul {\hat \mu} (n,t_r)$ at the
integer points
$x=n\in\bbZ$. Since the
$t_r$-dependence of
$\ul {\hat \mu}$ plays no role for this argument, we
ignore this dependence
in the following result.

\begin{lemma} \lb{dubn}
Abbreviate ${\ul A} =(A_1,\dots,A_\g)= {\ul
A}_{P_{\infty_-}}(P_{\infty_+})$ and $\ul {\hat
\mu} (n_0) = \ul {\hat \mu}^{(0)}$, with
$\ul {\hat \mu} (n)$ defined in
\eqref{ddt2}. Consider the
Dubrovin-type system
\begin{subequations} \lb{dubdis}
\begin{align}
\frac{\partial}{\partial x}\mu_j (x) &=
\frac{2y(\hat \mu_j(x))}{\prod_{\ell
\neq j}^\g
(\mu_j(x)-\mu_{\ell}(x))} \sum_{m=1}^\g
\Phi_{\g -m}^{(j)} (\ul \mu (x))
\sum_{n=1}^\g c_{m,n}^{-1}
A_n, \quad j=1,\dots,\g, \lb{dubdisa} \\
\ul {\hat \mu} (n_0) &= \ul {\hat \mu}^{(0)},
\lb{dubdisb}
\end{align}
\end{subequations}
with $c_{m,n}$ defined in \eqref{b6}. Denote the solution
of \eqref{dubdis} by
$\ul {\hat \mu}_0 (x)$. Then $\ul {\hat \mu} (n)$
coincides with $\ul {\hat
\mu}_0 (x)$ at integer
values $x=n\in\bbZ$, that is,
\begin{equation}
\ul {\hat \mu} (n) = \ul {\hat \mu}_0 (n),
\quad n\in\bbZ.
\end{equation}
\end{lemma}
\begin{proof}
First we recall the well-known result (see, e.g.,
\cite{BGHT98}, Sect. 3)
\begin{equation}
{\ul \alpha}_{P_0} (\calD_{\ul {\hat \mu} (n)}) -
{\ul \alpha}_{P_0}
(\calD_{\ul {\hat \mu} (n_0)})
=(n-n_0) \ul A_{P_{\infty_-}}(P_{\infty_+}) =(n-n_0)\ul A.
\end{equation}
In order to complete the proof we only need to stablish
that $\ul {\hat
\mu}_0 (x)$ satisfies
\begin{equation}
{\ul \alpha}_{P_0} (\calD_{\ul {\hat \mu}_0 (x)}) -
{\ul \alpha}_{P_0}
(\calD_{\ul {\hat \mu}_0
(x_0)})  =(x-x_0) \ul A_{P_{\infty_-}}(P_{\infty_+})
=(x-x_0)\ul A,
\end{equation}
that is, we need to show
\begin{equation}
\frac{\partial}{\partial x} {\ul \alpha}_{P_0}
(\calD_{\ul {\hat \mu}_0
(x)}) = \ul A. \lb{da}
\end{equation}
But equation \eqref{da} follows immediately from
\eqref{dubt} (identifying
$t_r$ with $x$ and
ignoring its $n$-dependence), \eqref{assump},
\eqref{mt} (multiplied by $i$), and
\eqref{dubdis}.
\end{proof}

Thus, the solution $\ul {\hat \mu}_0 (x)$ of \eqref{dubdis}
provides a
continuous interpolation for
$\ul {\hat \mu} (n)$. In fact, it was our attempt to
prove a result like Lemma
\ref{dubn} which led us to reconsider Dubrovin equations
and ultimately
resulted in Theorem
\ref{theorem1}, the explicit connection between
$\Phi_k^{(j)}(\ul \mu)$ and the polynomials
$\hatt F_r (z,x)$ defining the
hierarchy in question.

Toda systems as integrable discretizations of continuous systems are
also studied in \cite{Al91}.
%
%
\section{Theta function representations} \lb{s6}
The fundamental problem in the construction of
algebro-geometric solutions
of soliton
hierarchies is the following. Fix $\g\in\bbN$, and pick
a stationary
solution of
the $\g$th  equation in the hierarchy. Then solve the
$r$th time-dependent
equation (for any $r\in\bbN$) with the given stationary
solution as initial
datum and express
the solution in terms of the Riemann theta function
associated with
$\calK_\g$. In the KdV case
this procedure has been pioneered by Its and Matveev
in their celebrated
paper \cite{IM75}.
Subsequently, the algebro-geometric approach to integrable
equations was
developed in papes by
Date, Dubrovin, Krichever, Matveev, Novikov, Tanaka, and
others \cite{DT76a},
\cite{DT76b}, \cite{Du77},
\cite{Du81}, \cite{Du83}, \cite{DMN76}, \cite{Kr76},
\cite{Kr77a},
\cite{Kr77b}. We  briefly recall
the results for the hierarchies studied in this paper.
Detailed discussions
for the KdV, AKNS, and
Toda hierarchies as well as the sG equation and other
completely integrable
systems can be found,
for instance, in \cite{BBEIM94}, Chs.\ 3, 4, 6,
\cite{BGHT98}, \cite{Di91},
Sect.\ 12.4, \cite{DKN90},
\cite{GH97},
\cite{GH98}, \cite{GR96}, \cite{GRT96}, \cite{GK90},
\cite{It81},
\cite{MA81}, \cite{Ma86},
Sect.\ 4.4,
\cite{Mc79}, \cite{Me87}, \cite{MM79}, \cite{NMPZ84},
\cite{Pr85},
\cite{SW85}, \cite{To89}, Ch.\ 4
and the references therein. (Without explicitly repeating
this in each case
below, we exclude
collision points for $\mu_j (x,t_r)$, that is, we will
always assume
$\mu_j \neq \mu_{j'}$ for $j \neq j'$.)

\textbf{The KdV hierarchy.} Let $V^{(0)}(x)$ be a
stationary solution of
the $\g$th KdV equation
associated with divisor $\calD_{\ul {{\hat \mu}}^{(0)}(x)}$.
We seek the
$r$th KdV flow
\begin{equation}
\KdV_r(V)=0,  \quad V(x,t_{0,r}) = V^{(0)}(x),
\quad x\in\bbR.  \lb{6.1}
\end{equation}
The solution is given by the Its--Matveev formula
\cite{IM75}
\begin{equation}
V(x,t_r)=\Lambda-
2\partial_x^2 \ln\big(\theta({\ul z}(\ul {\hat \mu}
(x,t_r)), \quad
(x,t_r)\in\bbR^2,\lb{kdvth}
\end{equation}
where
\begin{equation}
{\ul {\hat \mu}} (x,t_{0,r}) = \ul {\hat \mu}^{(0)} (x)
\end{equation}
and
\begin{equation}
{\ul z}(\ul {\hat \mu} (x,t_r)) =
{\ul \alpha}_{P_0}(\calD_{\ul {\hat
\mu}(x,t_r)})
- \ul {\hatt A}_{P_0}(P_{\infty}) + \ul {\Xi}_{P_0}, \lb{kz}
\end{equation}
with $\ul {\Xi}_{P_0}$ the vector of Riemann constants
and $\Lambda$ a
$\calK_\g$-dependent constant.

\textbf{The AKNS hierarchy.}
Let $(p^{(0)}(x),q^{(0)}(x))$ be a stationary solution
of the $\g$th AKNS
equation associated with
divisors $\calD_{\ul {{\hat \mu}}^{(0)}(x)}$ and
$\calD_{\ul {{\hat
\nu}}^{(0)}(x)}$. We want to
construct the $r$th AKNS flow
\begin{equation}
\AKNS_r(p,q)=0,  \quad (p(x,t_{0,r}),q(x,t_{0,r}))=
(p^{(0)}(x),q^{(0)}(x)),
\quad x \in
\bbR.  \lb{6.4}
\end{equation}
The solution reads
\begin{subequations} \lb{aknsth}
\begin{align}
&p(x,t_r)=p(x_0,t_{0,r})
\f{\tht(\uz_-(\ul {\hat \nu}(x_0,t_{0,r})))}
{\tht(\uz_+(\ul {\hat \nu}(x_0,t_{0,r})))}
\f{\tht(\uz_+(\ul {\hat \nu}(x,t_r)))}
{\tht(\uz_-(\ul {\hat \nu}(x,t_r)))} \times
\lb{aknsthetap} \\
&\quad \quad\quad \quad\times
\exp(-2i(x-x_0)e_0-2i(t_r-t_{0,r})\tilde e_r),  \no \\
&q(x,t_r)=q(x_0,t_{0,r})
\f{\tht(\uz_+(\ul {\hat \mu}(x_0,t_{0,r})))}
{\tht(\uz_-(\ul {\hat \mu}(x_0,t_{0,r})))}
\f{\tht(\uz_-(\ul {\hat \mu}(x,t_r)))}
{\tht(\uz_+(\ul {\hat \mu}(x,t_r)))}\times  \lb{aknsthetaq} \\
&\quad \quad\quad \quad\times
\exp(2i(x-x_0)e_0+2i(t_r-t_{0,r})\tilde e_r), \no \\
& \hspace*{7cm} (x,t_r)\in\bbR^2, \no
\end{align}
\end{subequations}
where $\tilde e_r$ is a $\calK_\N$-dependent constant, and
\begin{equation}
{\ul {\hat \mu}} (x,t_{0,r}) = \ul {\hat \mu}^{(0)} (x),
\quad {\ul {\hat
\nu}} (x,t_{0,r}) = \ul
{\hat \nu}^{(0)} (x)
\end{equation}
and
\begin{equation}
\uz_\pm(\ul Q)=\ul\alpha_{P_0}(\calD_{\ul Q}) -
\ul A_{P_0}(P_{\infty_\pm})
+ {\ul\Xi}_{P_0},
\quad \ul Q=(Q_1,\dots,Q_\n), \lb{az}
\end{equation}
with $e_0$, $\tilde e_r$ are $\calK_\g$-dependent constants.

\textbf{The Toda hierarchy.} Let $(a^{(0)}, b^{(0)})$  be a
stationary solution of the $\g$th Toda equation associated
with divisor
$\calD_{\ul {{\hat
\mu}}^{(0)}(n)}$. We are interested in the
solution of the $\TL_r$ flow
\begin{equation}
\TL_r (a, b)=0, \quad (a(n,t_{0,r}), b(n,t_{0,r}))=
(a^\bze(n), b^\bze(n)),
\, \, n\in\bbZ.
\lb{3.6.6}
\end{equation}
Here the solution equals
\begin{subequations} \lb{todath}
\begin{align}
a(n,t_r) & =
\tilde a \left(\frac{\theta (\uz_+
(\ul {\hat \mu}(n-1,t_r)))\theta (\uz_+
(\ul {\hat \mu}(n+1,
t_r)))}{\theta (\uz_+
(\ul {\hat \mu}(n,t_r)))^2}\right)^{1/2},
\lb{todatha}\\
\begin{split}
b(n,t_r) & = \sum_{j=1}^\g \lambda_j
-\frac12 \sum_{m=0}^{2\g+1} E_m
-\sum_{j=1}^\g c_j (\g) \dfrac{\pa}{\pa w_j}
\ln \left(\dfrac{\theta (\uw
+\uz_+ (\ul {\hat \mu}(n,t_r)))}{\theta(\uw +
\uz_+ (\ul {\hat \mu}(n-1,t_r)))} \right) \Bigg|_{\uw =
\uzero}, \lb{todathb} \\
& \hspace*{7cm} (n,t_r) \in\bbZ\times \bbR,
\end{split}
\end{align}
\end{subequations}
where
\begin{equation}
{\ul {\hat \mu}} (n,t_{0,r}) = \ul {\hat \mu}^{(0)} (n)
\end{equation}
and
\begin{equation}
{\ul z}_+ (\ul {\hat \mu}(n,t_r))=
{\ul \alpha}_{P_0}(\calD_{\ul {\hat
\mu}(n,t_r)}) - {\ul
A}_{P_0}(P_{\infty_+})  + {\ul \Xi}_{P_0}, \lb{tz}
\end{equation}
with $\tilde a \neq 0$ a $\calK_\g$-dependent constant.

\textbf {The sGmKdV hierarchy.} Let $u^{(0)}(x)$ be the
solution of the $\N$th
stationary sGmKdV equation, that is,
\begin{equation}
g_{\N-1,x}-i(\beta e^{iu^{(0)}}-\alpha e^{-iu^{(0)}})=0,
\quad \N\in\bbN, \lb{4.1}
\end{equation}
subject to the constraints
\begin{equation}
\alpha \beta = \prod_{j=1}^{2\N}E_j, \quad E_0=0. \lb{4.1a}
\end{equation}
Let $r\in\bbN_0$. We are seeking the solution  $u$ of
$\sG_r(u)=0$ with $u(x,t_{0,r})=u^{(0)}(x)$. Here the
solution reads
\begin{align}
u(x,t_r)&=u(x_0,t_{r}) \lb{sgth}\\
&\quad +2i\ln\left(
\f{ \theta(z (\hat{\ul\mu}(x,t_r))+\ul\Delta) \theta
(z(\hat{\ul\mu}(x_0,t_{0,r})))}
{\theta (z(\hat{\ul\mu}(x_0,t_{0,r}))+\ul\Delta) \theta
(z(\hat{\ul\mu}(x,t_r))) }
\exp \big(- i e_0(x-x_0)\big) \right),\no
\end{align}
where $P_0=(0,0)$, $\ul \Delta$ is a half-period,
\begin{equation}
\ul \Delta = \ul A_{P_0}(\Pinf), \lb{3.31d}
\end{equation}
$e_0$ a $\calK_\N$-dependent constant, and
\begin{equation}
z(\ul Q)  =\ul\alpha_{P_0}(\calD_{\ul Q})
-\ul A_{P_0}(\Pinf)
+\ul\Xi_{P_0}.
\end{equation}
The linear equivalence of
$\calD_{\Pinf \hat {\ul \mu}(x,t_r)}$ and
$\calD_{Q_0 \hat {\ul \nu}(x,t_r)}$, that is,
\begin{equation}
\ul\alpha_{P_0}(\calD_{\hat{\ul\nu}(x,t_r)})=
\ul\alpha_{P_0}(\calD_{\hat{\ul\mu}(x,t_r)})+ \ul
\Delta, \lb{4.34p}
\end{equation}
shows that
\begin{equation}
z (\hat{\ul\nu}(x,t_r)) =z(\hat{\ul\mu}(x,t_{r}))+
\ul \Delta.
\end{equation}

In each of the theta function representations of this
section one should
keep in mind that the
change of coordinates effected by the Abel map straightens
out all
Dirichlet and Neumann flows\fnref{fn8}
on the Jacobi variety $J(\calK_\g)$ of $\calK_\g$, that is,
$z(\ul {\hat
\mu} (x,t_r))$ and
$z(\ul {\hat \nu} (x,t_r))$ in \eqref{kdvth},
\eqref{aknsth}, \eqref{todath} and  \eqref{sgth} are
linear in $x$
(respectively, $n$) and $t_r$.

%
%
\section{Examples}  \lb{s7}
\setcounter{theorem}{0}

\textbf{KdV and sG.} Pick $E_0=0$ and $E_1,\dots,
E_{2\g}\in\bbC$, $E_m
\neq E_{m'}$ for $m \neq m'$
and solve
\begin{subequations} \lb{mu}
\begin{align}
&\frac{\partial}{\partial x}\mu_{j}(x,t_\g)=-2i\,
\frac{y(\hat \mu_j(x,t_\g))}
{\prod_{\ell\neq j}^\g(\mu_j(x,t_\g)-\mu_\ell(x,t_\g))},
\lb{mux} \\
&\frac{\partial}{\partial t_\g}\mu_{j}(x,t_\g)=
\frac{\partial}{\partial
x}\mu_{j}(x,t_\g)\,
\frac1{16Q^{1/2}}\prod_{\ell\neq j}^\g\mu_\ell,
\quad  j=1,\dots,\g, \lb{mut}
\end{align}
\end{subequations}
with $Q=\prod_{m=1}^{2\g} E_m$ and
$R_{2\g+1}(z)=z\prod_{m=1}^{2\g}(z-E_m)$. Define
\begin{align}
u(x,t_\g)&=i\ln (Q^{-1/2} \prod_{j=1}^\g \mu_j(x,t_\g)), \\
V(x,t_\g)&=\sum_{m=0}^{2\g}E_m-2\sum_{j=1}^\g \mu_j(x,t_\g).
\end{align}
Then  $u$ and $V$ satisfy the sG equation and
$\g$th KdV equation,
respectively, that is,
\begin{equation}
 4 u_{x,t_\g}=\sin(u),  \quad \KdV_\g(V)=0
\end{equation}
for the following choice of $\tilde c_\ell$,
\begin{equation}
\tilde c_0=1,\quad \tilde c_1=\frac{(-1)^{\g-1}}
{16Q^{1/2}}-c_1(\ul E), \quad
\tilde c_{\ell} = -\sum_{p=0}^{\ell-1}\tilde c_p c_{\ell-p}(\ul E),
\quad
\ell=2,\dots,\g.
\end{equation}
The isomorphism between algebro-geometric $\KdV_\g$ and sG equations
is of course well-known and has been discussed, for instance, in
\cite{AA87}, \cite{AA87a}.

\textbf{AKNS and Tl.}
Pick $E_0,\dots, E_{2\g+1}\in\bbC$, $E_m \neq E_{m'}$
for $m \neq m'$ and solve
\begin{subequations} \lb{muakns}
\begin{align}
&\frac{\partial}{\partial x}\mu_{j}(x,t_r)=-2i\,
\frac{y(\hat \mu_j(x,t_r))}
{\prod_{\ell\neq j}^\g(\mu_j(x,t_r)-\mu_\ell(x,t_r))},
\lb{muaknsx} \\
&\frac{\partial}{\partial t_r}\mu_{j}(x,t_r)=
-\frac{\partial}{\partial
x}\mu_{j}(x,t_r)\,
\sum_{n=(r-\g)\maxi 0}^r d_n \Phi_n^{(j)}(\ul\mu),
\quad  j=1,\dots,\g \lb{muaknst}
\end{align}
\end{subequations}
and
\begin{subequations} \lb{nuakns}
\begin{align}
&\frac{\partial}{\partial x}\nu_{j}(x,t_r)=-2i\,
\frac{y(\hat \nu_j(x,t_r))}
{\prod_{\ell\neq j}^\g(\nu_j(x,t_r)-
\nu_\ell(x,t_r))}, \lb{nuaknsx} \\
&\frac{\partial}{\partial t_r}\nu_{j}(x,t_r)=
-\frac{\partial}{\partial
x}\nu_{j}(x,t_r)\,
\sum_{n=(r-\g)\maxi 0}^r d_n \Phi_n^{(j)}(\ul\nu),
\quad  j=1,\dots,\g,
\lb{nuaknst}
\end{align}
\end{subequations}
where  $R_{2\g+2}(z)=\prod_{m=0}^{2\g+1}(z-E_m)$
and $d_n\in\bbC$. Define
\begin{subequations} \lb{trakns}
\begin{align}
\frac{p_x(x,t_r)}{p(x,t_r)}&=i\sum_{m=0}^{2\g+1}E_m
-2i\sum_{j=1}^\g\nu_j(x,t_r), \lb{traknsp} \\
\frac{q_x(x,t_r)}{q(x,t_r)}&=-i\sum_{m=0}^{2\g+1}E_m
+2i\sum_{j=1}^\g\mu_j(x,t_r) \lb{traknsq}
\end{align}
\end{subequations}
and
\begin{subequations} \lb{trtoda}
\begin{align}
a(n,t_r)^2=&\frac12\sum_{j=1}^\g \frac{y(\hat \mu_j(n,t_r))}
{\prod_{k\neq j}^\g(\mu_j(n,t_r)-\mu_k(n,t_r))} \no \\
&-\frac14 b(n,t_r)^2-\frac14\sum_{j=1}^\g
\mu_j(n,t_r)^2+\frac18\sum_{m=0}^{2\g+1}E_m^2,
\lb{trtodaa} \\
b(n,t_r)=&-\frac12\sum_{m=0}^{2\g+1}E_m+
\sum_{j=1}^\g\mu_j(n,t_r).
\lb{trtodab}
\end{align}
\end{subequations}

Then $(p,q)$ and $(a,b)$ satisfy the $r$th AKNS equation
and the $r$th Toda lattice (Tl) equation, respectively,
that is,
\begin{equation}
\AKNS_r(p,q)=0, \quad \TL_r(a,b)=0 \lb{at}
\end{equation}
for the same choice of $c_\ell$, $\ell=1,\dots,r$ in
both equations
\eqref{at} (depending on the
choice of $d_n$ in \eqref{muaknst}, \eqref{nuaknst}).

\begin{remark}\lb{remark7.1} These examples provide 
interesting connections between
the $\KdV_\g$ and sG
equation (where $N$
is even and $0\in\{E_m\}_{m=0,\dots,N}$), and AKNS and
Toda hierarchies (where $N$ is odd), respectively,
and illustrate the
fundamental role of the
Dubrovin equations as the common underlying principle
for hierarchies of
soliton equations. In
particular, our approach establishes an isomorphism
between the classes of
algebro-geometric
solutions of these pairs of integrable systems. Indeed, once the
hyperelliptic curve $\calK_\g$ is fixed, algebro-geometric solutions
of the $\KdV_\g$ and sG (respectively, algebro-geometric solutions of
the $r$th AKNS and $r$th Tl equation) are just certain symmetric
functions (i.e., ``trace'' relations) of the solutions 
$\mu_1(x,t_\g),\dots,\mu_\g(x,t_\g)$ (resp.\ 
$\mu_1(x,t_r),\dots,\mu_\g(x,t_r), \nu_1(x,t_r),\dots,\nu_\g(x,t_r)$)
of the corresponding Dubrovin equations on $\calK_\g$.  Analagous
considerations apply to the nonlinear Schr\"odinger equation and the
(continuum) Heisenberg chain (see, e.g., \cite{El90}).  The
interesting problem which types of symmetric functions of
$\mu_1,\dots,\mu_\g$ (i.e., which types of ``trace'' formulas)
actually lead to completely integrable hierarchies is currently under
investigation. 
\end{remark}

%
%
\appendix
\section{Lagrange interpolation formulas} \lb{A}
\renewcommand{\theequation}{A.\arabic{equation}}
\renewcommand{\thetheorem}{A.\arabic{theorem}}
\setcounter{theorem}{0}
\setcounter{equation}{0}
In the following we suppress the $(x,t_r)$-dependence
as it will
be of no importance in this appendix.

Fix $\g\in\bbN$ and recall that
\begin{equation}
F_\g(z)=\prod_{j=1}^\g (z-\mu_j), \label{A1a}
\end{equation}
which implies that ($F_\g^\prime=
\partial F_\g/\partial z$)
\begin{equation}
F_\g^\prime(\mu_k)=
\prod_{\substack{j=1\\ j\neq k}}^\g (\mu_k-\mu_j).
\label{A1b}
\end{equation}
The general form of Lagrange's interpolation theorem
then reads  as
follows. (For convenience of the
reader we supply its proof even though the result is
well-known.)
\begin{theorem}\label{theoremA1}
Assume that $\mu_1,\dots,\mu_\g$ are $\g$ distinct
complex numbers. Then
\begin{multline}
\sum_{j=1}^\g
\f{\mu_j^{m-1}}{F_\g^\prime(\mu_j)}\Phi_k^{(j)}(\ul\mu)
=\delta_{m,\g-k}-\Psi_{k+1}(\ul\mu)\delta_{m,\g+1},
\\ m=1,\dots, \g+1,\quad k=0,\dots, \g-1. \label{A1c}
\end{multline}
\end{theorem}
\begin{proof}
Let $C_R$ be a circle with center at the origin and
radius $R$ that
contains the
zeros $\mu_j$ of the polynomial $F_\g$ and which is
oriented clockwise.
Cauchy's theorem then
yields
\begin{multline}
\f1{2\pi i}\oint_{C_R}d\zeta\,
\f{\zeta^{m-1}}{F_\g(\zeta)(\zeta-z)}=\f{z^{m-1}}{F_\g(z)}+
\sum_{k=1}^\g\f{\mu_k
^{m-1}}{F_\g^\prime(\mu_j)(\mu_j-z)},
\\  z\neq \mu_1,\dots, \mu_\g,
\quad m=1,\dots,\g+1.\label{A1d}
\end{multline}
However, by letting $R\to\infty$ we infer that
\begin{equation}
\f1{2\pi i}\oint_{C_R}d\zeta \, \f{\zeta^{m-1}}
{F_\g(\zeta)(\zeta-z)}
=\lim_{R\to\infty}\f{R^{m-1}}{F_\g(R)}=\delta_{m,\g+1},
\quad m=1,\dots,\g+1,
\label{A1e}
\end{equation}
which implies
\begin{equation}
z^{m-1}-\sum_{k=1}^\g\f{\mu_k^{m-1}
F_\g(z)}{F_\g^\prime(\mu_j)(z-\mu_j)}=F_\g(z)\delta_{m,\g+1}.
\label{A2}
\end{equation}
Using the symmetric functions $\Psi_j$ we may write
\begin{equation}
F_\g(z)=\sum_{j=0}^\g z^{\g-j}\Psi_j(\ul\mu) \label{A3}
\end{equation}
and
\begin{equation}
\f{F_\g(z)}{z-\mu_j}=
\sum_{k=0}^{\g-1}z^{\g-1-k}\Phi_k^{(j)}(\ul\mu).
\label{A4}
\end{equation}
Expanding both sides of equation \eqref{A2} in powers
in $z$, using
\eqref{A3} on the right-hand
side and \eqref{A4} on the left-hand side, proves
\eqref{A1c}.
\end{proof}

The simplest Lagrange interpolation formula reads in
the case $k=0$,
\begin{equation}
\sum_{j=1}^\g \f{\mu_j^{m-1}}{F_\g^\prime(\mu_j)}=
\delta_{m,\g}, \quad
m=1,\dots,\g. \label{A5}
\end{equation}

For  use in the main text we finally observe the following
equalities.  Adding
\eqref{A3} to $\mu_j$ times \eqref{A4} we find
\begin{equation}
F_\g(z)+\mu_j\f{F_g(z)}{z-\mu_j}=
\sum_{k=0}^{\g-1}z^{\g-k-1}(\Psi_{k+1}+
\mu_j\Phi_{k}^{(j)})+z^\g.\label{A6}
\end{equation}
However, we also have
\begin{equation}
F_\g(z)+\mu_j\f{F_\g(z)}{z-\mu_j}=z\f{F_\g(z)}{z-\mu_j}=
\sum_{k=0}^{\g-1}z^{\g-k-1}\Phi_{k+1}^{(j)}+z^\g,
\label{A7}
\end{equation}
using \eqref{A4} and recalling $\Phi_\g^{(j)}=0$.  Thus
we conclude
\begin{equation}
\Psi_{k+1}(\ul\mu)+\mu_j\Phi_{k}^{(j)}(\ul\mu)=
\Phi_{k+1}^{(j)}(\ul\mu),
\quad k=0,\dots,\g-1.
\label{A8}
\end{equation}
Finally, we will show
\begin{equation}
\sum_{\ell=0}^k \mu_j^\ell \Psi_{k-\ell}(\ul\mu)=
\Phi_{k}^{(j)}(\ul\mu), \quad
k=0,\dots,\g,\label{A9}
\end{equation}
by induction. Equation \eqref{A9} clearly holds for
$k=0$; next assume that
\begin{equation}
\sum_{\ell=0}^{k-1} \mu_j^\ell \Psi_{k-1-\ell}=
\Phi_{k-1}^{(j)}\label{A10}
\end{equation}
holds.  Then
\begin{align}
\sum_{\ell=0}^k \mu_j^\ell \Psi_{k-\ell}
&=\Psi_{k}+\mu_j\sum_{\ell=1}^k \mu_j^{\ell-1}
\Psi_{k-\ell} \\
&=\Psi_{k}+\mu_j\sum_{\ell=0}^{k-1} \mu_j^{\ell}
\Psi_{k-1-\ell}
=\Psi_{k}+\mu_j\Phi_{k-1}^{(j)}=\Phi_{k}^{(j)},\notag
\end{align}
using first the induction hypothesis and then \eqref{A8}.

{\bf Acknowledgments.} H.H.\ is indebted to the Department
of Mathematics at
the University of
Missouri, Columbia for the great hospitality
extended to him during his sabbatical 1996--97 when
major parts of this work were done.

\end{document}